# Galaxy Classification Using Transfer Learning and Ensemble of CNNs with Multiple Colour Spaces

Yevonnael Andrew

A thesis submitted in partial fulfilment of the requirements of
Liverpool John Moores University for the degree of
Master of Science in Artificial Intelligence and Machine Learning

June 2022

# ABSTRACT

Big data is now a norm in astronomy. The growth of astronomical images makes it a very suitable domain for computer science research. It is common for astronomer uses the morphological of galaxies to classify galaxies into categories. This practice was first applied systematically by Hubble (1936). When the data is small in size, the classification process could be easily done by small teams or individuals. However, the exponential growth of data collected by modern telescopes made it impossible to rely on an expert to classify every single galaxy image.

In December 2013, Winton Capital and Galaxy Zoo, together with the Kaggle team, created a Galaxy Challenge, where participants were asked to create a model to classify galaxies into categories. Since then, researchers worldwide have often used the Kaggle Galaxy Zoo dataset.

This research will focus on investigating how colour space transformation could affect classification accuracy, and we will investigate whether the CNN architecture will affect this too. For this research, we will consider multiple colour spaces (RGB, XYZ, LAB, etc) and multiple CNN architectures (VGG, ResNet, DenseNet, Xception, etc). We will use a pre-trained model and weights. However, most of the pre-trained model was trained on a natural RGB image, so we will investigate the performance in predicting the transformed (non-RGB) non-natural image (astronomical images).

We test our hypothesis by first testing individual networks using RGB and transformed colour spaces. We also test multiple ensemble configurations of different networks and colour spaces. We did a minimal hyperparameter search to ensure we obtained an optimum result. Our experimental results show that using transformed colour spaces on individual networks produced a higher validation accuracy. Ensembles of network and colour spaces further increase the validation accuracy.

Finally, this research aims to validate the usefulness of colour space transformation for astronomical images. This research will also become a benchmark that is useful for future research.

## LIST OF TABLES

# LIST OF FIGURES

## LIST OF ABBREVIATIONS

| | |
|---|---|
| BN | Batch Normalization |
| CIE | *Comission International de L'Éclairage* |
| CIFAR | Canadian Institute for Advanced Research |
| CNN | Convolutional Neural Network |
| FIRST | Faint Images of the Radio Sky at Twenty-cm |
| GPU | Graphics Processing Unit |
| GZ2 | Galaxy Zoo 2 |
| HSL | Hue, Saturation, Lightness (colour space) |
| HSV | Hue, Saturation, Value (colour space) |
| ILSVRC | ImageNet Large Scale Visual Recognition Challenge |
| LSTM | Long Short Term Memory |
| MLP | Multi Layer Perceptron |
| NTSC | National Television Standards Committee |
| NVSS | NRAO VLA Sky Survey |
| RGB | Red, Green, Blue (colour space) |
| SDSS | Sloan Digital Sky Survey |
| SECAM | *Séquentiel Couleur à Mémoire* |
| SGD | Stochastic Gradient Descent |
| VGGNet | Visual Geometry Group Deep Convolutional Networks |

**TABLE OF CONTENTS**

# CHAPTER 1

# INTRODUCTION

## 1.1 Background

Big data is now a norm in astronomy due to a shared culture of cooperation and regulations, in which the collective data from those telescopes are made available online (Feigelson and Babu, 2012). SDSS, which began operating in 2000 (Gunn *et al.*, 2006), now produces about 200 GB of data every night (Feigelson and Babu, 2012).

Fortunately, the growth of astronomical images happens simultaneously with the growth of computing power. Astronomy arguably is a perfect domain for computer science research because it pushes the boundaries of existing analysis (Kremer *et al.*, 2017). This makes sense because as data increases, analysis becomes more difficult. Common astronomical images found in our daily life are the colourised RGB version. Most original datasets of astronomical images do not correspond to the wavelength range that is sensitive to the human eye (Rector *et al.*, 2005). The way of astronomy image creation could affect how the image is interpreted. Hence, computer science researchers need to develop techniques that can address the nature of astronomical images, both in terms of their volumes and their uniqueness.

It is common for astronomer uses the morphological of galaxies to classify galaxies into categories. This practice was first applied systematically by Hubble (1936). Small teams or individuals could quickly process the classification process when the data is small in size. However, the exponential growth of data collected by modern telescopes made it impossible to rely on an expert to classify every single galaxy image.

To solve this, a crowdsourced project Galaxy Zoo project was launched (Lintott *et al.*, 2008), which invites volunteers to assist in classification. The project was very successful because, in approximately 175 days, more than 100,000 volunteers were involved in assisting classification (Lintott *et al.*, 2011).

The success of ConvNets in the ImageNet competition (2012) brought a revolution in computer vision; since then, ConvNets was becoming a dominant approach for almost all image-related tasks (LeCun, Bengio and Hinton, 2015). The architecture that won the ImageNet competition (Krizhevsky, Sutskever and Hinton, 2012) is now known as AlexNet.

Shortly after the success of the ImageNet competition, ConvNet also began to be widely used in astronomical images. The availability of datasets collected from the Galaxy Zoo project makes it a perfect situation to boost the usage of ConvNets in astronomy. In December 2013, Winton Capital and Galaxy Zoo, together with the Kaggle team, created a Galaxy Challenge, where participants were asked to create a model to classify galaxies into categories. Since then, researchers worldwide have often used the Kaggle Galaxy Zoo dataset.

There is a lot of possible ConvNet application in astronomy. For classification alone, current usage of CNN are galaxy classification (Dieleman, Willett and Dambre, 2015), solar radio spectrum classification (Chen *et al.*, 2017), sunspot group classification (Tang *et al.*, 2021), variable stars classification (Szklenár *et al.*, 2020), pulsar candidate classification (Wang *et al.*, 2019).

CNN can also be combined with other networks for a particular task. For example, the combination of CNN and LSTM can be used to classify transient radio frequency inference (Czech, Mishra and Inggs, 2018).

A digital image can be represented in different colour spaces (RGB, HSV, LAB, etc.). However, the most common colour space used in deep learning is RGB. The galaxy images from Kaggle Galaxy Challenge are also in RGB format. We are interested in how different colours can affect the model's performance. This research will consider all colour spaces and network architecture combinations on astronomical images. We will also compare the performance if we train the model from scratch and use the pre-trained model.

## 1.2 Problem Statement and Related Research

Kaggle Galaxy Challenge asked participants to create a model to classify galaxies. Dieleman, Willett and Dambre (2015) developed a deep neural network model which exploits rotational and translational symmetry. Dieleman's winning solution in the Kaggle competition required simple ensembling by averaging over 17 network variants and 60 transformations. The network variants differ in their number of dense layers, filter size configuration, activation function on the dense layer, and the number of filters.

Dieleman's winning solution incorporates the ensembling of networks. Ensembling is often used to increase the accuracy of the results. However, ensemble in deep learning is still not well studied. Most of the deep learning literature focuses on the design of the network, and most of them only applies naïve ensembling (Ju, Bibaut and van der Laan, 2018). The authors further test multiple ensembling methods on the CIFAR-10 dataset. Super Learner (Van der Laan, Polley and Hubbard, 2007) yield the highest accuracy over other ensembling methods. Super Learner finds the best weights adaptively without human intervention, which means we can include all the weak learners in the library.

Ensembling multiple networks has increased accuracy in numerous image-related tasks, including astronomical images. However, none of them studied in-depth ensembling multiple networks trained in different colour spaces.

ColorNet (Gowda and Yuan, 2019) shows that transforming RGB colour into different colour spaces can significantly affect classification accuracy. Using a simple convolutional network on the CIFAR-10 dataset on multiple colour spaces (RGB, HSV, LAB, etc) shows that LAB colour spaces yield the highest accuracy. The experiment also shows that each class has different accuracy for each colour space, which means there is no perfect correlation between colour spaces. Finally, ColorNet proposed an ensemble of DenseNet based models with seven colour spaces to obtain high classification accuracy.

In other fields, some works exist that tried to do colour space transformations before doing further analysis. For example, in medical images, conversion to CIE Lab coluor space for dysplastic nuclei segmentation achieves the best-averaged accuracy (dos Santos *et al.*, 2020). CIE Lab also yields the highest classification accuracy in histological image classification (Velastegui and Pedersen, 2021). The authors also further show that despite CIE Lab yielding the highest overall accuracy, some classes are more accurate when they are represented in other colour spaces. FusionNet (Guo *et al.*, 2020) used YIQ colour space for multi-modal medical image fusion. In the field of steganography, StegColNet (Gowda and Yuan, 2021) shows that the ensemble of the colour space model outperforms the recent state-of-the-art approach.

### 1.3 Research Questions

After conducting literature reviews, several research questions would like to be addressed by this research.

- How does the colour transformation affect the classification accuracy in the astronomical images?
- Would the network learn different representations when we feed them with different colour spaces?

- How does the usage of pre-trained models and training from scratch affect classification accuracy?
- How does the usage of different base architecture for ensembling affect the classification accuracy?
- What are the best ensemble methods to combine multiple networks trained on different colour spaces in order to yield a higher accuracy?

### 1.4 Aim and Objectives

This research aims to understand the effect of colour transformation and ensembles of CNN trained on different colour transformations on the classification accuracy of astronomical images.

Based on the aim of this study, the research objectives are formulated as follows:

- To analyse the effect of colour transformation with respect to classification accuracy
- To compare the effect of different CNN architectures used to train images with different colour space
- To analyse whether using a pre-trained model suitable for astronomical images (considering most of the pre-trained model was trained on daily object images)
- To identify relevant research related to colour space transformations on CNN
- To identify relevant research on the usage of CNN in astronomical images
- To compare between the ensembling methods to improve classification accuracy
- Develop a new methodology so that they can combine the benefit of multiple colour spaces and ensembling

This research is intended to develop a novel methodology that can effectively combine networks trained on multiple colour spaces. We hypothesise that the combination of these networks will improve classification accuracy.

### 1.5 Significance of the Study

This research aims to contribute new pieces of knowledge to the deep learning and computer vision fields in the following ways:

- To validate the usefulness of colour space transformation to increase the classification accuracy of astronomical images
- To validate the usefulness of transfer learning for astronomical images

- A benchmark for the comparison of different CNN architecture on galaxy classification
- A benchmark of multiple ensembling methods on galaxy classification

### 1.6 Scope of the Study

Taking into account the time and resource constraints, we will limit the scope of the study as follows:

- This work only focuses on astronomical data, i.e., Kaggle Galaxy Challenge data.
- Other methods, such as statistical feature extraction, will not be considered here.

# CHAPTER 2

# LITERATURE REVIEW

## 2.1. Machine Learning in Astronomy

The earliest published work exploring the use of machine learning in astronomy was by Adorf and Meurs (1988), which used both supervised and unsupervised classification to classify the IRAS Point Source Catalog.

The earliest published work to automatically classify stars/galaxies using neural networks was by Odewahn *et al* (1992). The authors use both linear perceptron and multi-layer perceptron for the classification task. The classification is based on manually extracted features, e.g. diameter, ellipticity, average transmission, central transmission, gradients, etc.

The earliest review paper on neural network applications in astronomy was published by Miller (1993). The author identified several major areas of research: adaptive telescope optics, object classification, object matching, and detector event filtering. The state-of-the-art algorithm in this paper is the *multi-layer perceptrons employing back-propagation learning* (MLP) and *self-organising maps* (SOM).

The back-propagation algorithm is a learning procedure, in which the weights of connections are adjusted iteratively to minimise the difference between the true value and the predicted value (Rumelhart, Hinton and Williams, 1986). Years later, the back-propagation algorithm was applied to solve a real-world problem, recognising handwritten zip codes (LeCun *et al.*, 1989).

Neural networks and back-propagation were developed in the 1980s. The ability of the algorithms was also already demonstrated by LeCun in 1989. However, there was a fundamental problem in deep learning, which made researchers lose interest. By the late 1980s, it was known that traditional deep feedforward networks were hard to train by back-propagation (Schmidhuber, 2015). The reason is that deep neural networks suffer from a problem that is now famous as exploding and vanishing gradients (Hochreiter, 1991).

Due to the limitation of back-propagation of recurrent neural networks, i.e., exploding and vanishing gradients, an important concept called Long Short Term Memory (LSTM) was developed (Hochreiter and Schmidhuber, 1997). But, the LSTM breakthrough did little to fix the larger problem of neural networks and did not work very well; also, computers were not fast enough, algorithms were not smart enough, and people were not satisfied (Kurenkov, 2020).

In the 1990s, the enthusiasm and optimism on AI are at a low point. Thus, this period is often referred to as AI Winter, when the funding and interest in AI research were reduced (*AI Newsletter*, 2005). This situation was confirmed by work from LeCun *et al*, which compares learning algorithms for recognising handwritten digits (LeCun *et al.*, 1995). The paper compares classification algorithms developed at Bell Laboratories and elsewhere, i.e., Linear Classifier (Baseline), Nearest Neighbor Classifier (Baseline), Pairwise Linear Classifier, PCA and Polynomial Classifier, RBF Network, Large Fully Connected Multi-Layer Neural Network, LeNet 1, LeNet 4, LeNet 5, Boosted LeNet 4, Tangent Distance Classifier and Optimal Margin Classifier. Using the Optimal Margin Classifier (Boser, Guyon and Vapnik, 1992), a test error of 1.1% was reached. LeNet 4 has a test error of 1.1%, while the best algorithm is Boosted LeNet, with a test error of 0.7%. It shows that the Optimal Margin Classifier, now known famous as Support Vector Machine, worked better or the same compared to the neural networks.

A decision-tree-based classifier called Random Decision Forests (Ho, 1995) was developed, and the validity is demonstrated by experiments on recognising handwritten digits. Random Forests are proven to be very effective and come with a sound mathematical theory (Kurenkov, 2020). This also contributes to the AI winter.

## 2.2. Convolutional Neural Networks

Convolutional neural networks, which are sometimes referred to as ConvNets or CNNs, are a specialised kind of neural network that is suitable for data that has a grid-like topology, like images, which are usually thought of as a two-dimensional grid of pixels (Goodfellow, Bengio and Courville, 2016).

### 2.2.1 AlexNet

In 2012, there was a competition called *ImageNet Large Scale Visual Recognition Challenge*, which is now often referred to as *ImageNet* only. The winning solution achieved a top-5 error of 15.3%, which is substantially lower than the runner-up. This winning solution is now well known as AlexNet (Krizhevsky, Sutskever and Hinton, 2012).

AlexNet has eight layers – five convolutional and three fully-connected layers. To prevent overfitting, the network uses ReLUs non-linearity (Nair and Hinton, 2010) and dropout layer (Hinton *et al.*, 2012). While ReLU does not require input normalisation, applying normalisation after ReLU in certain layers successfully reduces error rates. This architecture also used overlapping pooling, which reduces the error rates compared with non-overlapping pooling.

To combat memory limitation, the network is trained across two GPUs, where each GPU has half of the neurons. The GPU communicates with each other on a certain layer, e.g., layer 3 take input from all kernel from two GPUs in layer 2; layer 4 only take input from the kernel of layer 3 within the same GPU.

AlexNet uses two data augmentation strategies. The first strategy is done by extracting random 224 x 224-pixel patches from 256 x 256 images, along with their horizontal reflections. This augmentation is done with little computation and is not stored in the disk. The second strategy is by altering RGB intensities, i.e., performing PCA on the RGB pixel values.

The model was trained using SGD (stochastic gradient descent) with batch size 128, the momentum of 0.9, and weigh decay 0.0005. The learning rate was initialised at 0.01, divided by 10 if the validation rate was not improving.

### 2.2.2 VGGNet

VGGNet is a class of neural networks that employs a very deep network for image recognition (Simonyan and Zisserman, 2014). This architecture won ImageNet Challenge 2014 competition - first place in the localisation track and second place in the classification track.

VGGNet uses a 3 x 3 receptive field, 1-pixel convolution stride, and spatial padding such that spatial resolution after convolution is preserved. Max-pooling window size is 2 x 2 pixels, with stride 2. The total pooling layer for each architecture is five. Thus, not every convolution is followed by pooling. The fully connected layers consist of two 4096 channels and one 1000 channels, with a soft-max layer as the final layer. All layers use ReLu for non-linearity. This paper also uses LRN normalisation (Krizhevsky, Sutskever and Hinton, 2012) in one of the architectures, but it does not improve the performance.

### 2.2.3 GoogLeNet

GoogLeNet, an Inception-based deep convolutional neural network with 22 layers, achieved a new state of the art in the ImageNet Challenge 2014 (Szegedy *et al.*, 2015).

A set of techniques were adopted to obtain higher performance:

- Ensembling 7 models independently with the same architecture, same initialisation, and same learning rate policies. The difference between models is in the sampling methodologies and random order of input image.
- Data augmentation involves resizing, cropping, and mirroring, which leads to 144 crops for each image.
- The softmax is the average over multiple crops and over all the individual classifiers. The authors tested that max-pooling over crops and averaging over the individual classifiers lead to inferior performance.

### 2.2.4 Xception

Xception (Chollet, 2017), "Extreme Inception," is a neural network architecture that uses a "depthwise separable convolution." Xception has a slightly better performance on the ImageNet dataset.

### 2.2.5 ResNet

ResNet (He *et al.*, 2016). The network's convolutional layers mostly have filters of 3 x 3. The number of filters is designed to preserve the time complexity per layer. Downsampling is done by convolutional layers of stride 2. The network ends with a global average pooling layer. Then we will insert "shortcut connections."

The input is processed by the following methods (Krizhevsky, Sutskever and Hinton, 2012; Simonyan and Zisserman, 2014). To reduce the *internal covariate shift*, Batch Normalization (Ioffe and Szegedy, 2015) is applied before activation and right after each convolution. The dropout layers are not needed because of the regularisation provided by BN. The model was trained using SGD (stochastic gradient descent) with batch size 256, momentum of 0.9, and weight decay of 0.0001. The learning rate was initialised at 0.01, which was divided by 10 when plateaus. The models trained for up to $60 \times 10^4$ iterations.

### 2.2.6 DenseNet

Dense Convolutional Network (Huang *et al.*, 2017) is a type of neural network architecture where each layer takes all preceding feature maps as input. This is the benefit of DenseNet, *collective knowledge* of preceding feature maps.

If $L$ denotes the number of layers, then Densenet has $\frac{L(L+1)}{2}$ connections. In DenseNet, each layer applies a non-linear transformation $H_l(\cdot)$, which is defined as three consecutive operations: Batch Normalisation, ReLU, and 3 x 3 convolution.

The network is divided into multiple densely connected *dense blocks*. The layer between the dense blocks is referred to as *transition layers*, which consists of batch normalisation, 1 x 1 convolution layer, and 2 x 2 average pooling layer.

## 2.3. Pre-trained Model and Transfer Learning

The term "transfer learning" can be traced to its earliest work by Stevo and Ante (1976, 2020). In machine learning and deep learning, the distributions of training and testing data are assumed to be the same. Thus, if there is a discrepancy between training and testing data, the model may not work well, and the model needs to be rebuilt from scratch (Pan and Yang, 2009). For example, a model trained to discriminate astronomical images captured by an old telescope may not work well to predict astronomical images from the newer telescope as they will have a different quality.

Fortunately, transfer learning allows us to use training and testing data that have different distributions and different tasks and domains (Pan and Yang, 2009). The use of transfer learning allows us not to train the model from scratch.

In computer vision, transfer learning is done by using a pre-trained model.

### 2.3.1 Transfer Learning in Astronomy

Transfer learning has been used in astronomy as well. There are two common types of transfer learning implementation in astronomy: transfer learning from one survey to another or transfer learning from ImageNet. The second one is interesting considering the pre-trained model usually was trained on the ImageNet dataset, which contains natural images like animals, cars, etc.

*Table 1: Literature review summary on transfer learning in astronomy*

| **Authors and Year** | **Objective/Purpose** | **Architecture/Methods Used** |
|---|---|---|
| (George, Shen and Huerta, 2018) | Glitch classification and clustering of gravitational waves | Inception, ResNet, VGG. The trained CNN also used as a feature extractor for clustering |
| (Ackermann *et al.*, 2018) | Galaxy merger detection | Xception |
| (Tang, Scaife and Leahy, 2019) | Radio galaxy classification | Transfer learning from different surveys (NVSS and FIRST) |
| (A. Khan *et al.*, 2019) | Galaxy classification | Xception. The model also used as a feature extractor for clustering |
| (Yang *et al.*, 2020) | Lunar impact crater identification and age estimation | ResNet101 as a feature extractor |
| (Awang Iskandar *et al.*, 2020) | Planetary nebulae classification | InceptionResNetV2, DenseNet201, MobileNetV2 |
| (Wei *et al.*, 2020) | Star cluster classification | VGG19-BN, ResNet18 |
| (Tanoglidis, Ćiprijanović and Drlica-Wagner, 2021) | Separating low surface brightness galaxies from artifacts | Transfer learning from different surveys (DES and HSC-SSP) |
| (Farrens *et al.*, 2022) | Blended sources identification | VGG |

Table 1 shows literature that uses transfer learning in the astronomy field. Except noted, all pre-trained models are using the ImageNet dataset. For example, Tang, Scaife, and Leahy (2019) use transfer learning from different surveys (NVSS and FIRST) for radio galaxy classification.

### 2.3.2 Pre-trained Model as a Feature Extractor

A pre-trained model can be used as a features extractor as an image representation (Sharif Razavian *et al.*, 2014). The authors use the first fully-connected layer as a feature vector of size 4096, which is further trained on linear SVM for a classification task.

This method is also known as *deep feature extraction*.

*Table 2: Literature review summary on feature extraction using a pre-trained model*

| **Authors and Year** | **Objective/Purpose** | **Architecture/Methods Used** |
|---|---|---|
| (Chaib *et al.*, 2017) | Remote sensing image classification | CaffeNet + VGG-VD16. Features are extracted from the first FC layer, which then combined using either of two options: addition or concatenation. |
| (Lopes and Valiati, 2017) | Tuberculosis detection | GoogleNet, VGG, Resnet. Three different proposals: 1) Simple CNN feature extraction, 2) Bag of CNN features, 3) Ensembles |
| (Rajaraman *et al.*, 2018) | Malaria parasite detection | AlexNet, VGG16, Xception, ResNet50, DenseNet121. Candidate layers of the network were tested to get the optimal layer. |
| (Varshni *et al.*, 2019) | Pneumonia detection | Xception, VGG, ResNet, and DenseNet are used as feature extractors, followed by RF, KNN, NB and SVM. Hyperparameter tuning is done in the SVM classifier. |
| (Saxena, Shukla and Gyanchandani, 2020) (Saxena, Shukla and Gyanchandani, 2020) | Breast cancer detection | Ten different pre-trained CNNs were investigated. Images are divided into non-overlapping patches. Linear SVM classifier used as final classifier. |
| (Barbhuiya, Karsh and Jain, 2021) | Sign language classification | Modified pre-trained Alexnet and VGG16 used as feature extractor, followed by SVM classifier. |

### 2.4. Colour Space

Light in the visible region of the electromagnetic spectrum that falls upon the human retina is what we know as colour (Poynton, 1997). The human retina has three types of colour photoreceptor cells. Thus, three numerical components are needed to describe a colour.

The first defined quantitative links between electromagnetic spectrums and perceived colour by humans are by the *Comission International de L'Éclairage* (Smith and Guild, 1931; CIE, 1932), which created two colour spaces: CIE 1931 RGB colour space and CIE 1931 XYZ colour space. CIE 1931 colour space model defines three parameters denoted as "X", "Y", and "Z", where Y is the luminance component, and additional X and Z components.

#### 2.4.1 Transformation to XYZ

The standardised transformation equations recommended by CIE (Fairman, Brill and Hemmendinger, 1997) are as follows:

$$\begin{bmatrix} X \\ Y \\ Z \end{bmatrix} = \begin{bmatrix} 0.49 & 0.31 & 0.2 \\ 0.17697 & 0.81240 & 0.01063 \\ 0 & 0.01 & 0.99 \end{bmatrix} \begin{bmatrix} R \\ G \\ B \end{bmatrix}$$

### 2.4.2 Transformation to HSL

HSL (hue, saturation, lightness) represents an image in terms of chromatic and achromatic information. The commonly used equations (Hanbury and Serra, 2002) are:

$$S = \begin{cases} 0 & \text{if } \max(R,G,B)=\min(R,G,B) \\ \dfrac{\max(R,G,B) - \min(R,G,B)}{\max(R,G,B) + \min(R,G,B)} & \text{if } L \leq 1/2 \\ \dfrac{\max(R,G,B) - \min(R,G,B)}{2 - [\max(R,G,B) + \min(R,G,B)]} & \text{otherwise} \end{cases}$$

The L component, which expresses the brightness, is expressed as:

$$L = \frac{\max(R,G,B) + \min(R,G,B)}{2}$$

The H component is expressed as:

$$H' = \begin{cases} \text{undefined} & \text{if } S=0 \\ \dfrac{G - B}{\max(R,G,B) - \min(R,G,B)} & \text{if } R=\max(R,G,B) \\ 2 + \dfrac{B - R}{\max(R,G,B) - \min(R,G,B)} & \text{if } G=\max(R,G,B) \\ 4 + \dfrac{R - G}{\max(R,G,B) - \min(R,G,B)} & \text{if } B=\max(R,G,B) \end{cases}$$

To get the final H value, H' is multiplied by 60°, $H = H' \times 60^o$.

### 2.4.3 Transformation to HSV

The H component of HSV is the same as the H component of HSL. The S and V components can be expressed as:

$$V = max(R,G,B)$$

$$S = \begin{cases} \dfrac{max(R,G,B) - min(R,G,B)}{max(R,G,B)} & \text{if } \max(R,G,B) \neq 0 \\ 0 & \text{otherwise} \end{cases}$$

### 2.4.4 Transformation to CIELAB

LAB colour space, also known as CIE 1976 L*a*b*, and CIELAB colour space. The conversion from RGB to LAB cannot be done straightforward, but we need to convert it into XYZ, followed by the conversion into LAB using the following equations (Schanda, 2007):

$$L^* = 116f\left(\frac{Y}{Y_n}\right) - 16$$

$$a^* = 500\left[f\left(\frac{X}{X_n}\right) - f\left(\frac{Y}{Y_n}\right)\right]$$

$$b^* = 200\left[f\left(\frac{Y}{Y_n}\right) - f\left(\frac{Z}{Z_n}\right)\right]$$

where, being t = $t = \frac{X}{X_n}, \frac{Y}{Y_n}, or \frac{Z}{Z_n}$:

$$f(t) = \begin{cases} \sqrt[3]{t} & \text{if } t > \left(\frac{24}{116}\right)^3 \\ \left(\frac{841}{108}\right)(t) + \frac{16}{116} & \text{if } t \leq \left(\frac{24}{116}\right)^3 \end{cases}$$

**2.4.5 Transformation to CIELUV**

CIELUV is also known as the CIE 1976 L*, u*, v* colour space. The L* of CIELUIV is the same as that of the CIELAB. The other coordinates are defined as follows (Schanda, 2007):

$$u^* = 13L^*(u' - u'_n)$$

and

$$v^* = 13L^*(v' - v'_n)$$

Following the CIE 1960 UCS Diagram recommendation, the equations for u' and v' are as follow (Schanda, 2007):

$$u' = \frac{4X}{X + 15Y + 3Z}$$

and

$$v' = \frac{9Y}{X + 15Y + 3Z}$$

The value of $u'_n$ and $v'_n$ can be taken from the *white reference's* table (Poynton, 2012). For example, in CIE III C, $u'_n = 0.2009, v'_n = 0.4609$; in CIE III $D_{65}$, $u'_n = 0.1978, v'_n = 0.4683$.

**2.4.6 Transformation to YUV**

YUV is created from RGB, where Y is created from weighted values of R, G, and B as a measure of luminance. U and V are computed as differences between the calculated Y and B and R.

Derived from BT.470-6 (*1998*), the formula to convert RGB to YUV is as follows (equasys GmbH):

$$\begin{bmatrix} Y \\ U \\ V \end{bmatrix} = \begin{bmatrix} 0.299 & 0.587 & 0.114 \\ -0.147 & -0.289 & 0.436 \\ 0.615 & -0.515 & -0.100 \end{bmatrix} \begin{bmatrix} R \\ G \\ B \end{bmatrix}$$

The RGB value should be in the interval of [0,1].

### 2.4.7 Transformation to YCbCr

YCbCr, sometimes written as $YC_BC_R$. From full-scale 8-bit RGB, YCbCr can be computed using the following equations (Itu-t, 2011):

$$Y = 0.299 * \mathrm{R} + 0.587 * G + 0.114 * B$$

$$C_B = 128 + \frac{(-0.299 * \mathrm{R} - 0.587 * G + 0.886 * B)}{1.772}$$

$$C_R = 128 + \frac{(0.701 * \mathrm{R} - 0.587 * G - 0.114 * B)}{1.402}$$

### 2.4.8 Transformation to YDbDr

YDbDr, sometimes written $YD_BD_R$, is a colour space used in the SECAM TV system, which is used in France and some Eastern European countries. To convert RGB into YDbDr can be done using the following equations (Shi and Sun, 2019):

$$\begin{bmatrix} Y \\ Db \\ Dr \end{bmatrix} = \begin{bmatrix} 0.299 & 0.587 & 0.114 \\ -0.450 & -0.883 & 1.333 \\ -1.333 & 1.116 & -0.217 \end{bmatrix} \begin{bmatrix} R \\ G \\ B \end{bmatrix}$$

### 2.4.9 Transformation to YIQ

YIQ has been used in NTSC TV systems for years. To convert RGB into YIQ can be done using the following equations (Broesch, 2008; Shi and Sun, 2019):

$$\begin{bmatrix} Y \\ I \\ Q \end{bmatrix} = \begin{bmatrix} 0.299 & 0.587 & 0.114 \\ 0.596 & -0.275 & -0.321 \\ 0.212 & -0.523 & 0.311 \end{bmatrix} \begin{bmatrix} R \\ G \\ B \end{bmatrix}$$

### 2.4.10 Transformation to HED

HED (Haematoxylin-Eosin-DAB) is a special purpose colour space used in the medical field to analyse tissues (Ruifrok and Johnston, 2001), with the conversion equations as follow:

$$\begin{bmatrix} H \\ E \\ D \end{bmatrix} = \begin{bmatrix} 1.88 & -0.07 & -0.60 \\ -1.02 & -1.13 & -0.48 \\ -0.55 & -0.13 & 1.57 \end{bmatrix} \begin{bmatrix} R \\ G \\ B \end{bmatrix}$$

## 2.5. Color Space Transformation in CNN

This is a summary of the literature review for the use of colour space transformation in CNN. We only include research that involves transforming original colour space into other colour spaces, e.g., RGB to LAB.

*Table 3: Literature review summary on colour space transformation in CNN*

| **Author** | **Colour Space** | **Task Type** | **Field/Dataset** |
|---|---|---|---|
| (Rachmadi and Purnama, 2015) | RGB, LAB, XYZ, HSV | Classification | Vehicle images |
| (Kim and Ro, 2016) | RGB, LAB, YCbCr, HSV, YIQ, XYZ, RQCr, RIQ, YQCr | Face Recognition | Multi-PIE |
| (Atha and Jahanshahi, 2017) | RGB, YCbCr | Detection | Corrosion images |
| (Jafarbiglo, Danyali and Helfroush, 2018) | LAB | Classification | MITOS-ATYPIA-14 |
| (Gowda and Yuan, 2019) | RGB, HSV, LAB, YUV, YCbCr, YpBPr, YIQ, XYZ, HED, LCH, CMYK | Classification | CIFAR-10, CIFAR-100, CVHN, ImageNet |
| (M. A. Khan *et al.*, 2019) | LAB | Classification | Weizmann, KTH, UIUC, Muhavi, WVU |
| (Castro *et al.*, 2019) | RGB, HSV, LAB | Classification | Fruit dataset |
| (Li *et al.*, 2020) | RGB, HSV | Segmentation | Medical |
| (Mohammadi Lalabadi, Sadeghi and Mireei, 2020) | RGB, HSV, LAB | Classification | Fish dataset |
| (Gowda and Yuan, 2021) | RGB, HSV, LAB, YUV, YCbCr, YpBPr, YIQ, XYZ, HED, LCH, CMYK | Classification | Bossbane, BOWS2 |

Using LAB colour space for face recognition tasks results in higher accuracy than using RGB (Kim and Ro, 2016), 79.77% and 71.50%, respectively. The authors also propose "collaborative feature learning", a framework to aggregates features from multiple colour spaces, which further improves accuracy to 90.81%.

However, colour space transformation does not always bring a better result. For example, a publication on vehicle colour recognition (Rachmadi and Purnama, 2015) achieves slightly higher accuracy using RGB colour space compared to XYZ, LAB, and HSV – 0.9447, 0.9432, 0.9414, and 0.9372, respectively. The authors didn't address the stochastic factor of neural network – whether the accuracy differences were just a random chance, nor evaluate the use of different architecture on different colour spaces.

### 2.6. Deep Learning Ensemble

Ensemble learning combines several models to reduce generalisation error. If the models make independent errors, the ensemble will perform significantly better than its individual models (Goodfellow, Bengio and Courville, 2016).

Many machine learning competitions are won by the use of ensemble models or ensemble learning. Several reasons behind the success of ensemble learning are (Dietterich, 2000):

- Statistical reason. Ensemble reduces the risk arising from a single model and can help us find a good approximation to *f*.
- Computational reason. A single model may get stuck in local optima. An ensemble may start from different starting points to avoid getting stuck in the same position.
- Representational reason. In machine learning, sometimes, the function *f* cannot be represented by any hypothesis in a given space. By using an ensemble, the approximated function may expand beyond the given space.

#### 2.6.1 Unweighted model averaging

This approach is the most used approach in the literature. To get the final prediction, the outputs of learners are averaged. The averaging is performed either on the outputs of learners or by using the softmax function (Ju, Bibaut and van der Laan, 2018; Ganaie and Hu, 2021).

This simple averaging of six ResNet models with different depths is used as the winning solution for ILSVRC 2015 classification tasks (He *et al.*, 2016). VGGNet (Simonyan and Zisserman, 2014) also uses this ensemble method to get a lower test error. GoogleNet trained 7 versions of the same model, with different sampling methodologies and random input order, to create ensemble prediction (Szegedy *et al.*, 2015). However, this naïve averaging is not data-adaptive: it works well for networks that have comparable performance, and are sensitive to the excessively biased learners (Ju, Bibaut and van der Laan, 2018).

#### 2.6.2 Majority Voting

This approach is similar to unweighted averaging. Majority voting works by counting the number of all predicted labels from each individual learner. The final prediction depends on which label has the most votes. The majority voting is less sensitive to the excessively biased learners. (Ju, Bibaut and van der Laan, 2018)

# CHAPTER 3

# RESEARCH METHODOLOGY

## 3.1 Introduction

This chapter presents the methodology that will be used in this research. Section 3.2.1 talks about the dataset that will be used in this research. The section also talks about converting the original dataset into a format suitable for classification. The original images are in the RGB format and need to be converted into multiple colour spaces, which are discussed in section 3.2.2. To get a better generalisation, data augmentation will be discussed in section 3.2.3. The modelling approach will be discussed in section 3.2.4, followed by an evaluation in section 3.2.5. The proposed method will be discussed in section 3.3.

## 3.2 Research Methodology

This section discusses the methodology used in this research based on literature reviews in chapter 2. The methodology described is end-to-end, from dataset creation to ensemble models.

### 3.2.1 Dataset

The dataset used in this research was downloaded from Galaxy Zoo – The Galaxy Challenge. The dataset contains 61578 images in JPG and RGB format. This Galaxy Challenge vote fraction is a modified version of The Galaxy Zoo 2 project (Willett *et al.*, 2013). The Galaxy Zoo 2 (GZ2) has a total of 11 tasks and 37 possible responses.

The original challenge is a regression task, in which participants should predict the vote percentage, and the lowest RMSE determined the winner. The probability values in each set of responses are the likelihood of the galaxy falling in that category. The sum of all possible responses in one category is 1.0. For example, a galaxy had 75% of all users identify as smooth (Class 1.1), 15% as features/disk (Class 1.2), and 10% as a star/artefact. The total probability of Class 1 is 75% + 15% + 10% = 100%.

- Class 1.1 = 0.75
- Class 1.2 = 0.15
- Class 1.3 = 0.10

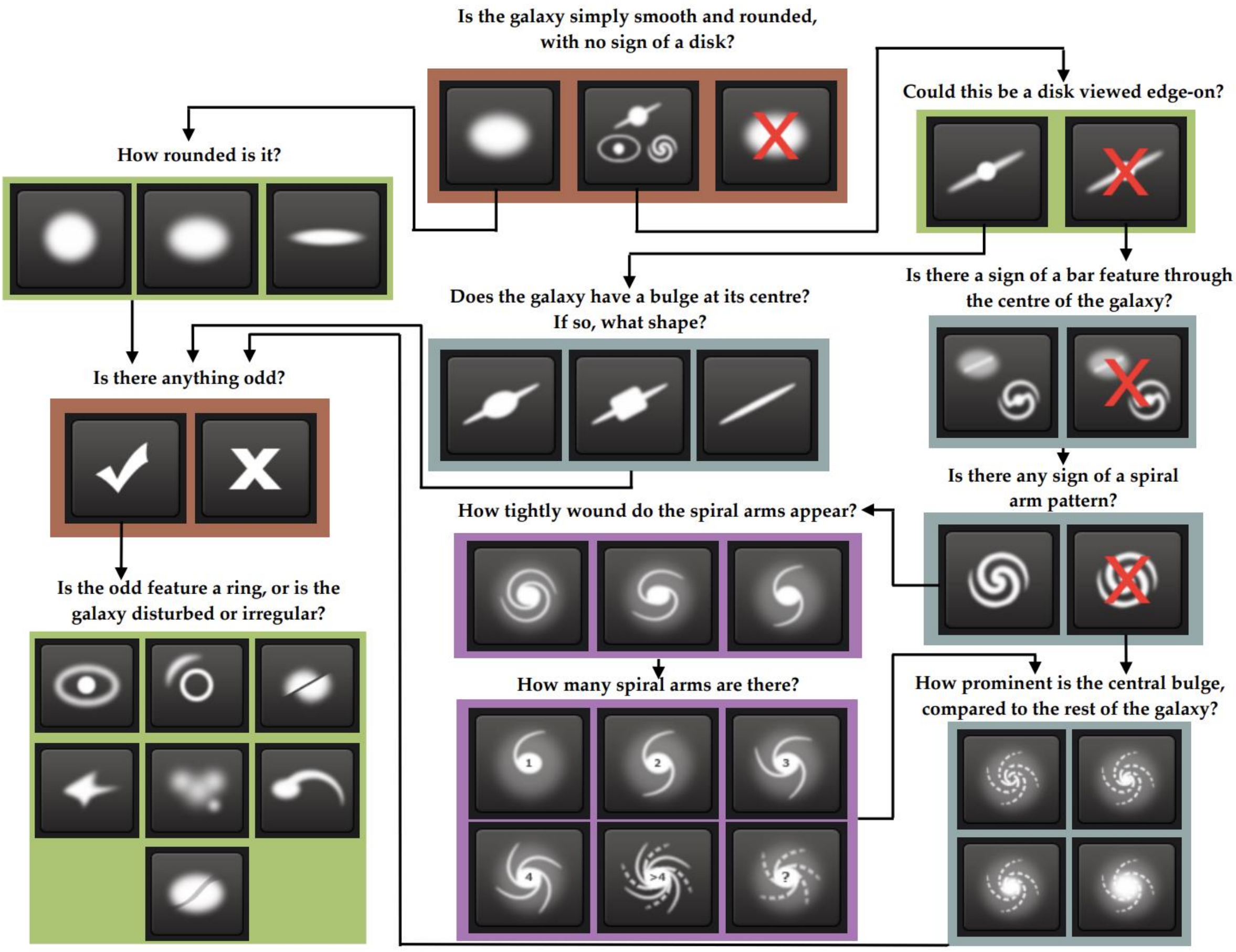

*Figure 1: Classification flowchart for Galaxy Zoo (Willett et al., 2013)*

Thus, to make it suitable for the classification task, we should convert them into classes. The conversion can be as simple as thresholding. For example, if we want to create a binary classification task, i.e., to predict whether the galaxy is smooth or features/disk, we can use a 50% vote as the threshold. This number should be determined carefully because we should incorporate statistics uncertainty.

*Table 4: Text representation of classification flowchart (Willett et al., 2013)*

| **Task** | **Question** | **Responses** | **Next** |
|---|---|---|---|
| 01 | Is the galaxy simply smooth and rounded, with no sign of disk? | smooth<br>features or disk<br>star or artefact | 07<br>02<br>end |
| 02 | Could this be a disk viewed edge-on? | yes<br>no | 09<br>03 |
| 03 | Is there a sign of a bar feature through the centre of the galaxy? | yes<br>no | 04<br>04 |
| 04 | Is there any sign of a spiral arm pattern? | yes<br>no | 10<br>05 |
| 05 | How prominent is the central bulge, compared with the rest of the galaxy? | no bulge<br>just noticeable<br>obvious<br>dominant | 06<br>06<br>06<br>06 |
| 06 | Is there anything odd? | yes<br>no | 08<br>end |
| 07 | How rounded is it? | completely round<br>in between<br>cigar-shaped | 06<br>06<br>06 |
| 08 | Is the odd feature a ring, or is the galaxy disturbed or irregular? | ring<br>lens or arc<br>disturbed<br>irregular<br>other<br>merger<br>dust lane | end<br>end<br>end<br>end<br>end<br>end<br>end |
| 09 | Does the galaxy have a bulge at its centre? If so, what shape? | rounded<br>boxy<br>no bulge | 06<br>06<br>06 |
| 10 | How tightly wound do the spiral arms appear? | tight<br>medium<br>loose | 11<br>11<br>11 |
| 11 | How many spiral arms are there? | 1<br>2<br>3<br>4<br>more than four<br>can't tell | 05<br>05<br>05<br>05<br>05<br>05 |

We will follow an appropriate threshold (Willett *et al.*, 2013) to select clean samples for this research. The final data will be represented by five classes, i.e., "edge-on", "spiral", "completely round smooth", "cigar-shaped smooth", and "in-between smooth".

*Table 5: Classification threshold (Willett* et al.*, 2013)*

| **Class Name** | **Tasks** | **Threshold** |
|---|---|---|
| Completely round smooth | T01<br>T07 | $f_{smooth} \geq 0.469$<br>$f_{completely\ round} \geq 0.50$ |
| In-between smooth | T01<br>T07 | $f_{smooth} \geq 0.469$<br>$f_{in-between} \geq 0.50$ |
| Cigar-shaped smooth | T01<br>T07 | $f_{smooth} \geq 0.469$<br>$f_{cigar-shaped} \geq 0.50$ |
| Edge-on | T01 | $f_{features/disk} \geq 0.430$ |
| | T02 | $f_{edge-on,yes} \geq 0.602$ |
| Spiral | T01<br>T02<br>T04 | $f_{features/disk} \geq 0.430$<br>$f_{edge-on,no} \geq 0.715$<br>$f_{spiral,yes} \geq 0.619$ |

The classes created from the given threshold are called *clean samples* because they are well-sampled.

### 3.2.2 Data Transformation

After obtaining the clean samples, the RGB images will be converted to multiple colour spaces using an open-source image-processing Python library, OpenCV and Scikit-image. The colour spaces available are HED, HSV, LAB, RGBCI E, XYZ, YCbCr, YDbDr, YIQ, YPbPr, and YUV. When we predict an image, a network that is trained on a specific colour space will only predict that specific colour space.

### 3.2.3 Data Augmentation

Data augmentation is a common practice in deep learning. We use data augmentation to make the network more robust to novel images, hence preventing overfitting and increasing the final accuracy (O'Gara and McGuinness, 2019). Following the paper results, we will apply data augmentation into our training process.

### 3.2.4 Modelling

There are a lot of readily available CNN architectures. We will test our transformed images with several architectures, including AlexNet (Krizhevsky, Sutskever and Hinton, 2012), VGG (Simonyan and Zisserman, 2014), Inception (Szegedy *et al.*, 2015), ResNet (He *et al.*, 2016), DenseNet (Huang *et al.*, 2017). DenseNet based ensembles are used in ColorNet (Gowda and Yuan, 2019).

We will consider two training scenarios:

- Train the entire model from scratch (using only the architecture) as a benchmark
- Using a pre-trained model as a base, freeze the convolutional layers and train only the fully-connected layers

According to the results that we obtained from transformed images trained on several CNN architectures, we will create an ensemble (Figure 1) to yield a higher classification accuracy.

The number of networks and what colour space we should use for each network are variables for this research.

### 3.2.5 Evaluation

We assume no class is more important than the other for galaxy classification. Thus, the model will be evaluated using accuracy.

$$accuracy = \frac{correct\ prediction}{total\ images} = \frac{TP + TN}{TP + FP + TN + FN}$$

## 3.3 Ensemble Method

Figure 1 shows the general workflow used in this research. In this section, multiple ensembles will be tested. Two types of ensembles will be tested. The first one is to ensemble the prediction made by each network by using simple averaging or a meta-learner. The second one is to extract feature vectors by dropping the fully-connected layers and combine them with feature vectors from other networks to be fed into a machine learning classifier.

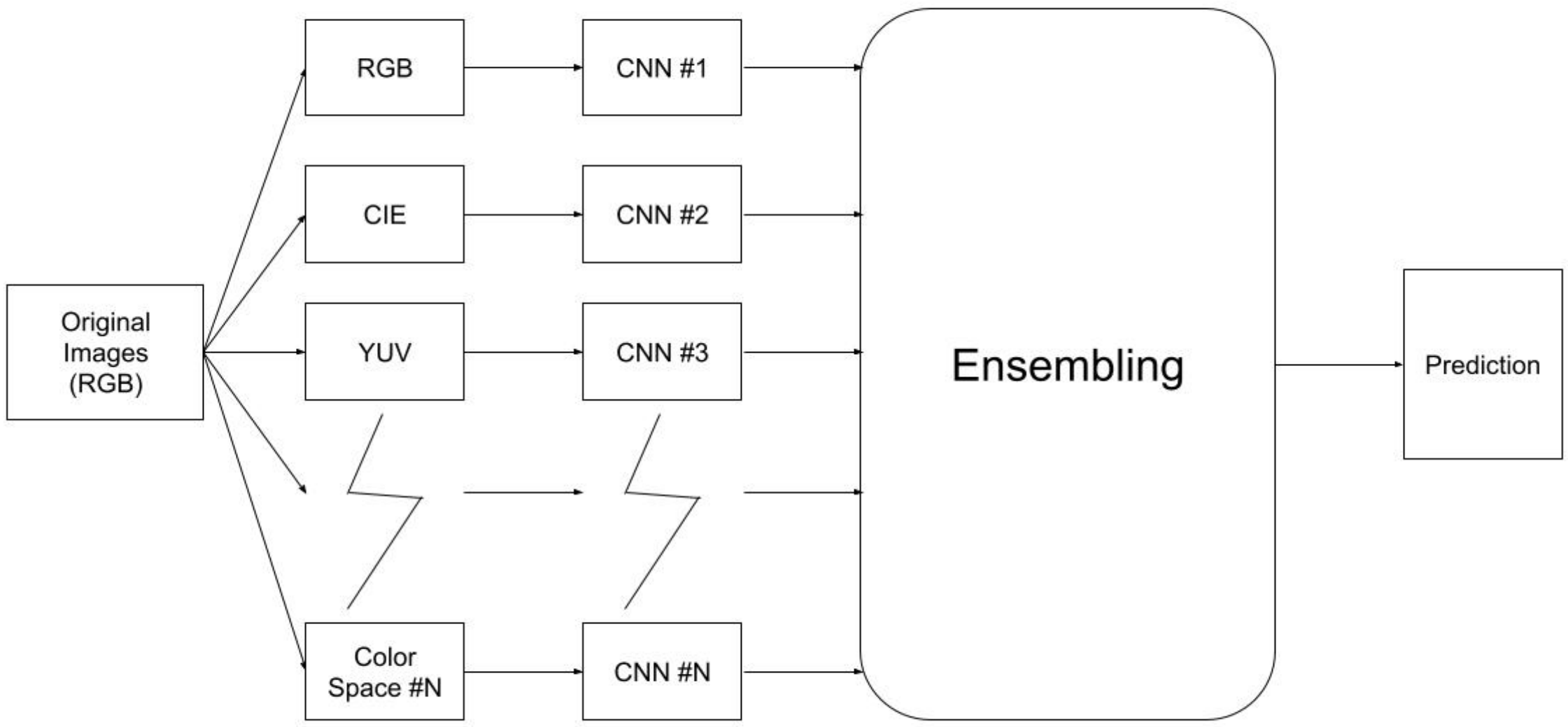

*Figure 2: Proposed Architecture*

## 3.4 Summary

All combinations in this research can be summarised as follows:

*Table 6: Combination of variables tested in this research*

| Variable Tested | Variants |
|---|---|
| Architecture | ResNet, VGG, DenseNet, Xception |
| Colorspaces | RGB, HED, HSV, LAB, XYZ, YCbCr, YDbDr, YIQ, YPbPr, YUV |
| Pre-trained and transfer learning | 1. Training from scratch (as a benchmark)<br>2. Using a pre-trained model as a feature extractor<br>3. Fine-tuning (train top classification later only) |
| Ensemble method | 1. Concatenate multiple networks just after flatten layer |

# CHAPTER 4

# IMPLEMENTATION

## 4.1 Introduction

This chapter describes the technical implementation of this research. Section 4.2. describes the dataset creation process, including label creation that is suitable for classification problems. Section 4.3 describes the colour space transformation. Section 4.4 describes the technical detail of network configurations.

## 4.2 Dataset Creation

This section describes the dataset creation that will be used for this research.

### 4.2.1 Image Resizing

The dataset used in this research was downloaded from Kaggle. The original image was in JPG format with the size of 424 x 424 x 3, where 424 x 424 is the width and height of the image, and 3 is three layers of RGB.

| **151220.jpg** | **157161.jpg** | **102422.jpg** |
| --- | --- | --- |
| 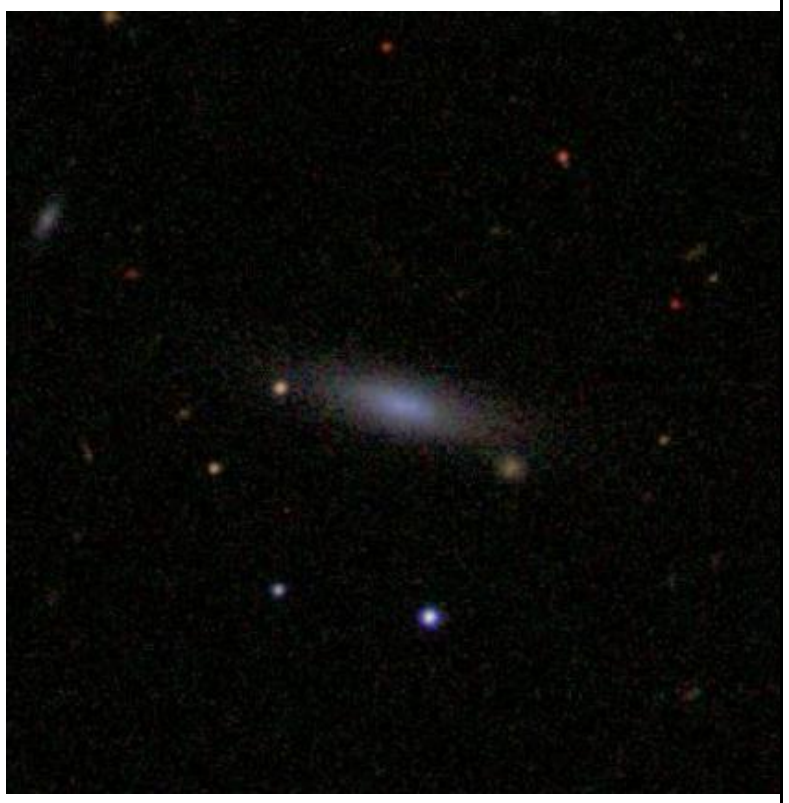 | 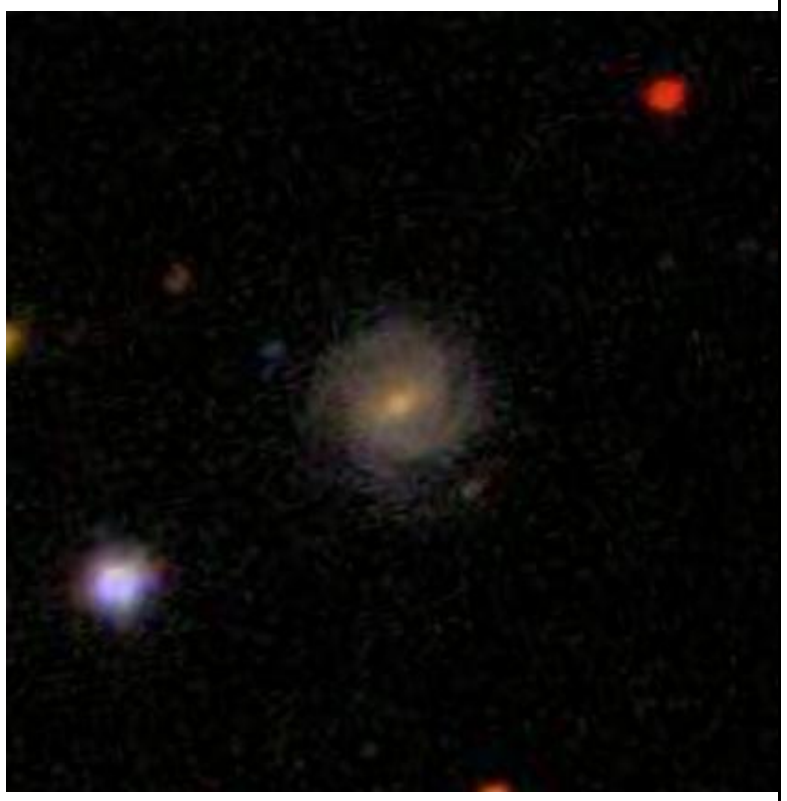 | 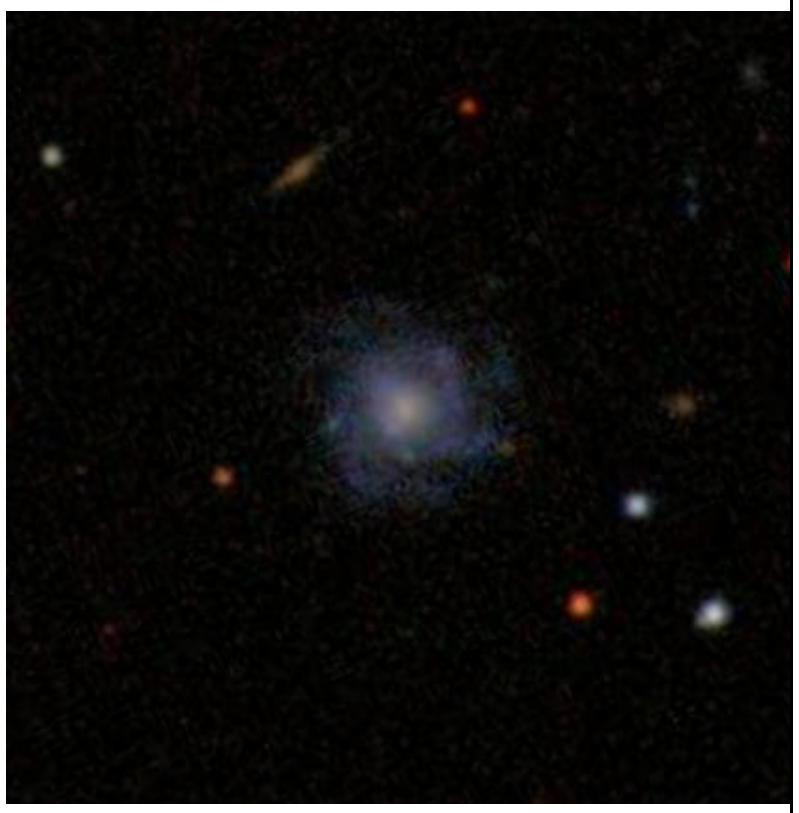 |

*Figure 3: Sample images*

Looking at Figure 3, it is clear that the most important part of the image is in the centre. The outer part of the image is either not helpful for galaxy classification or contains unrelated information that may be messing up the model and prediction results. Cropping images also help to speed up the training process, even though it is not the primary motivation here.

| 151220.jpg | 157161.jpg | 102422.jpg |
| --- | --- | --- |
| 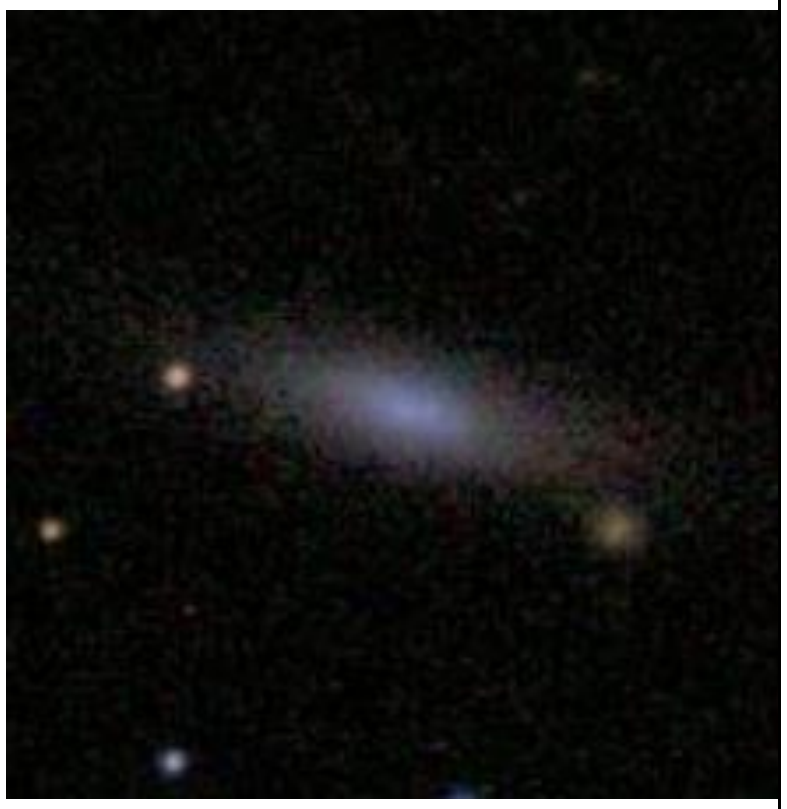 | 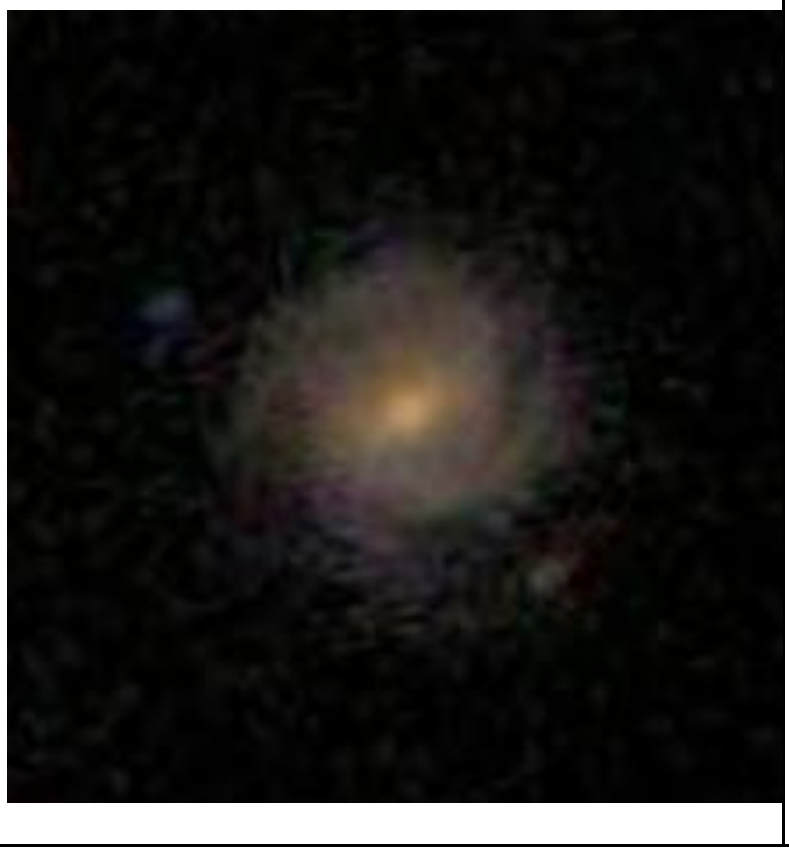 | 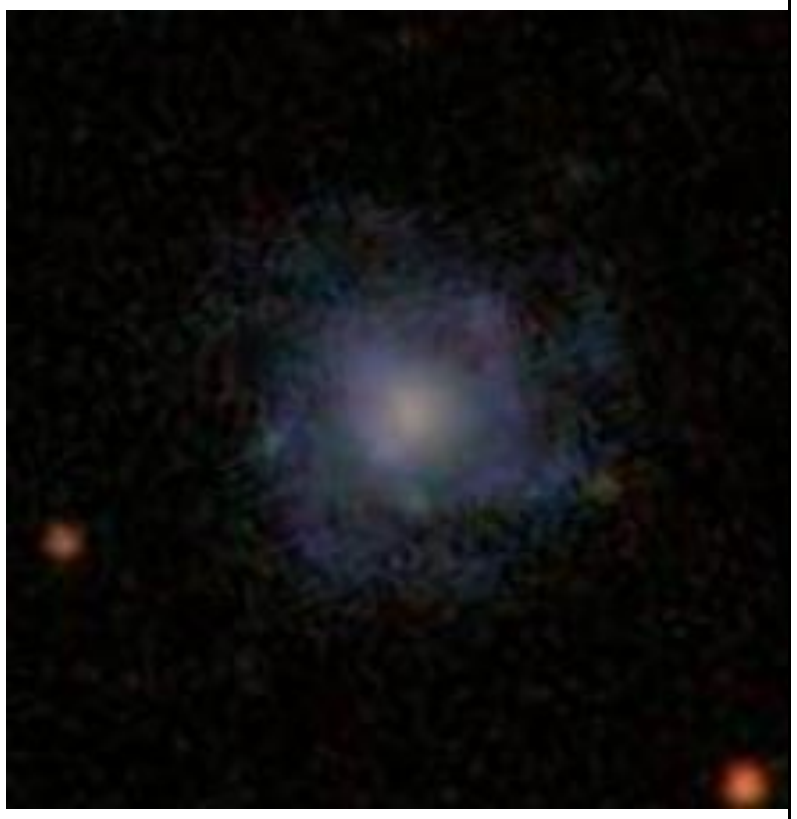 |

Figure 4: Cropped Images

We crop the outermost 100 pixels of the image. This left us with an image size of 224 x 224 x 3.

### 4.2.2 Label Creation for Classification

The original dataset was meant for regression problems. Thus, to make it suitable for this research, we should create a label that can be used for classification. The label creation rule and threshold are described in chapter 3.

| | | |
| --- | --- | --- |
| ***completelysmooth*** | 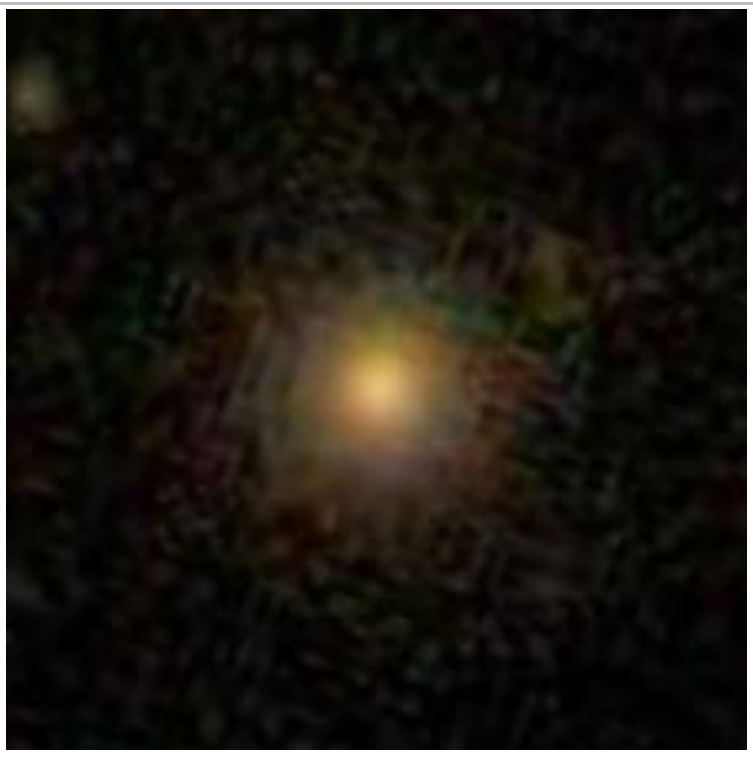 | 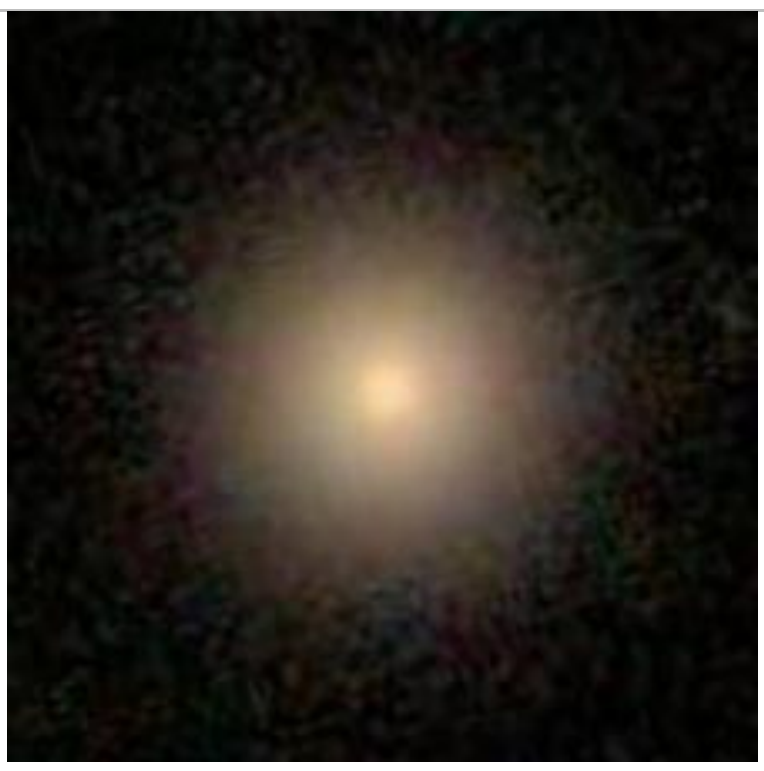 |
| ***inbetweensmooth*** | 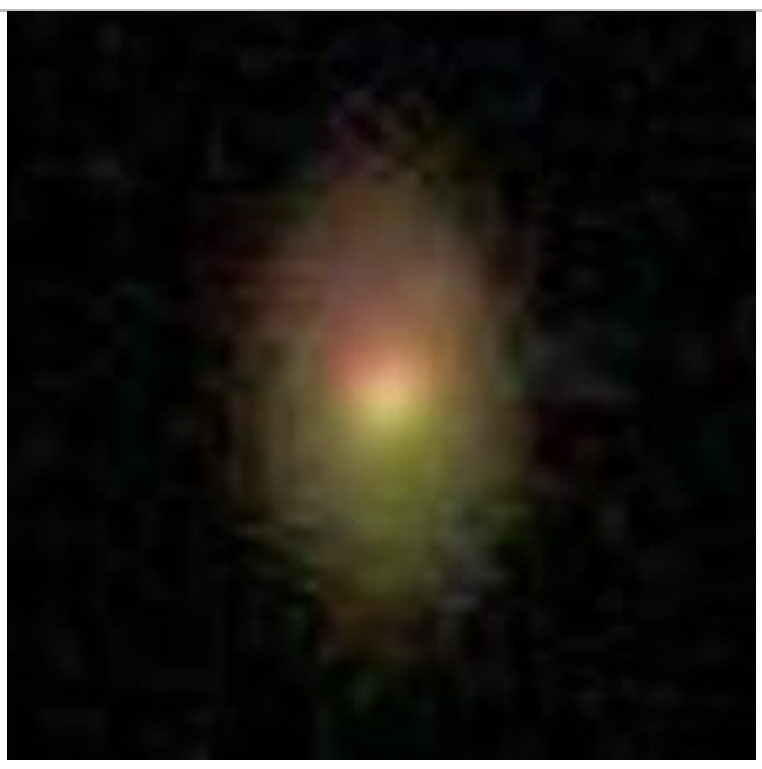 | 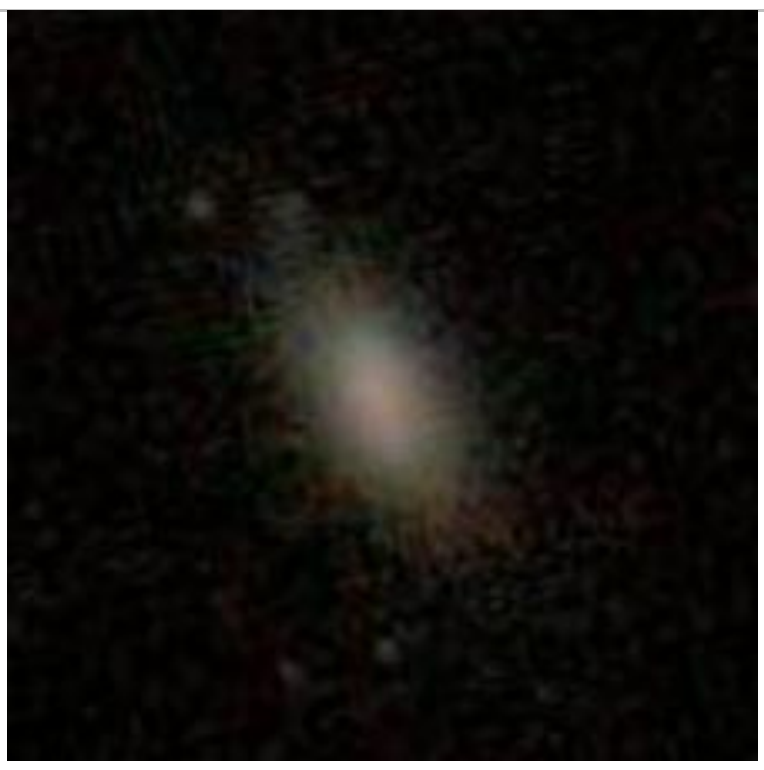 |

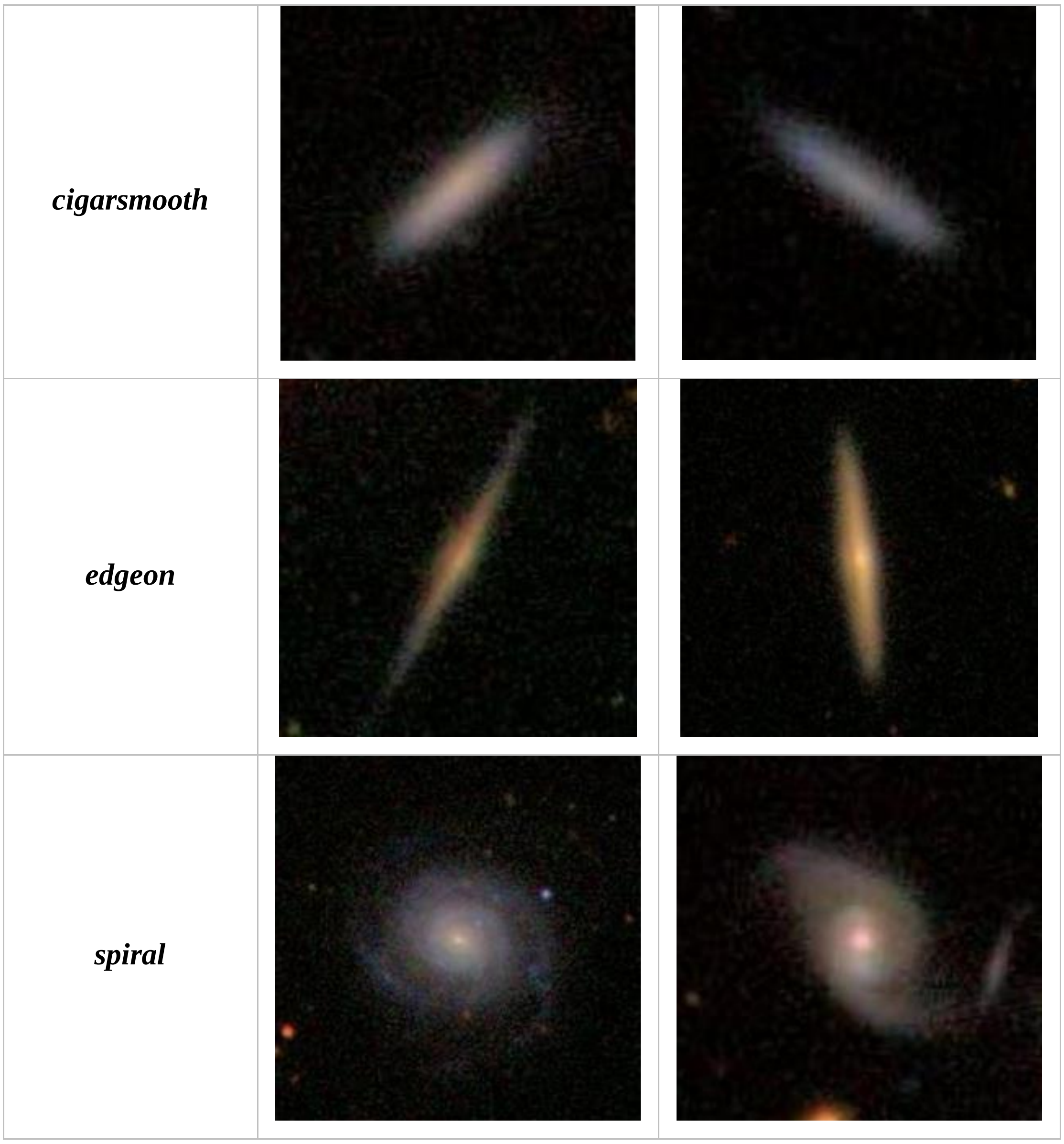

*Figure 5: Sample Images from Each Class*

The final labels consist of:

- *completelysmooth*: 8436 images
- *inbetweensmooth*: 8069 images
- *cigarsmooth:* 579 images
- *edgeon*: 3903 images
- *spiral*: 7806 images

### 4.3 Colour Space Conversion

The next step of the dataset creation is to convert the image to multiple colour spaces. Fortunately, TensorFlow has built-in colour space conversion for the following colour spaces:

LAB, XYZ, HSV, YUV, and YDbDr. The built-in colour space conversion is done by using tfio.experimental.color.rgb_to_lab(x).

For the rest of colour spaces (HLS, LUV, YCrCb, YIQ, HED), we need to manually do the conversion using either OpenCV or scikit-image and save it to disk for future usage.

### 4.4 Building Network

Building a network in deep learning is not a simple task. There are a lot of options, parameters, hyperparameters, and configurations to choose from. This further could also be affected by the type of data we use (natural images, medical images, etc.) and the type of task we would like to accomplish (classification, regression, segmentation, etc.). Hence, in this section, we test several commonly used values for every parameter and choose the one that produces the best result for our networks.

#### 4.4.1 Load Pre-trained Model Base Model and Weights

For this research, we used out-of-the-box pretrained model from TensorFlow. Architecture used in this research are:

- Xception
- VGG16, VGG19
- ResNet50, ResNet101, ResNet152
- DenseNet121, DenseNet169, DenseNet201

When loading the model, we set include_top = "False" to use the base model only. To load pre-trained weights, we set parameter weights = "imagenet". All images fed into networks will be resized to 128 x 128 x 3. We also set the base model to be non-trainable. The only trainable part in this research is the top layers.

#### 4.4.2 Stack Classification Layers

For top layers, which are the trainable part of the network, we should strive for a balance between the number of parameters and performance. We test four configurations of top layers:

- Configuration A: Dense(128)
- Configuration B: Dense(128) with Dropout(0.2)
- Configuration C: Dense(512) + Dense(128)
- Configuration D: Dense(512) + Dense(128) with Dropout(0.2)

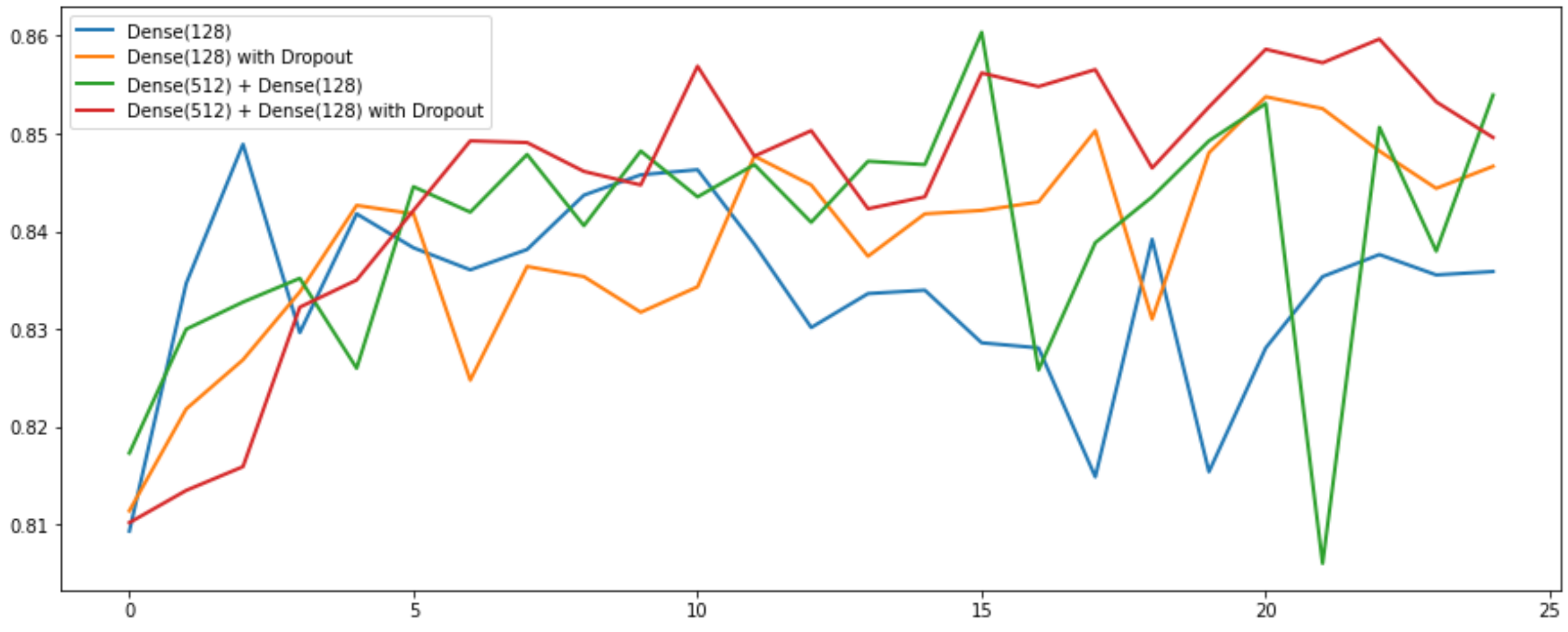

*Figure 6: Validation Accuracy of Multiple Top Layer Configurations*

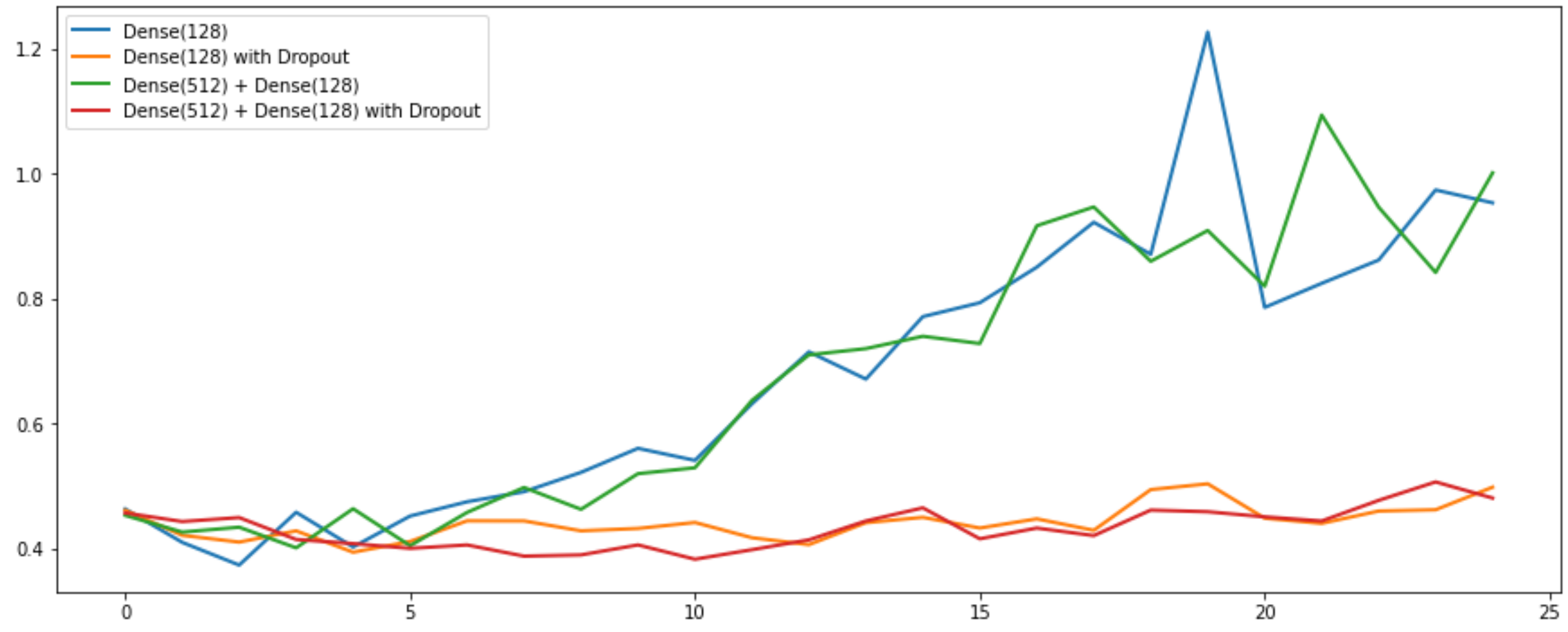

*Figure 7: Validation Loss of Multiple Top Layer Configurations*

Figure 4 shows that the validation loss of layers with Dropout (configuration B and D) are more stable than layers without Dropout (configuration A and C). While the validation loss of configurations B and D are similar, Figure 3 shows that configuration D has the highest accuracy. Thus, we will choose configuration D – Dense(512) + Dense(128) with Dropout(0.2) – for our top classification layers.

### 4.4.3 Configuring for Performance

To improve the training speed i.e., reduces step time, we implement *prefetching* and set the number of elements to prefetch using *AUTOTUNE*. We also implement *caching*, which will reuse the data cached for the next epoch. This configuration are chained and implemented using DATASET.cache().prefetch(buffer_size=AUTOTUNE).

### 4.4.4 Data Augmentation

Using data augmentation in the training pipeline is a common practice because it consistently improves final results. We benchmark two configurations in detail as follows:

- Configuration 1: Without data augmentation
- Configuration 2: With data augmentation
    - Random Horizontal Flip
    - Random Rotation with a factor of 0.2

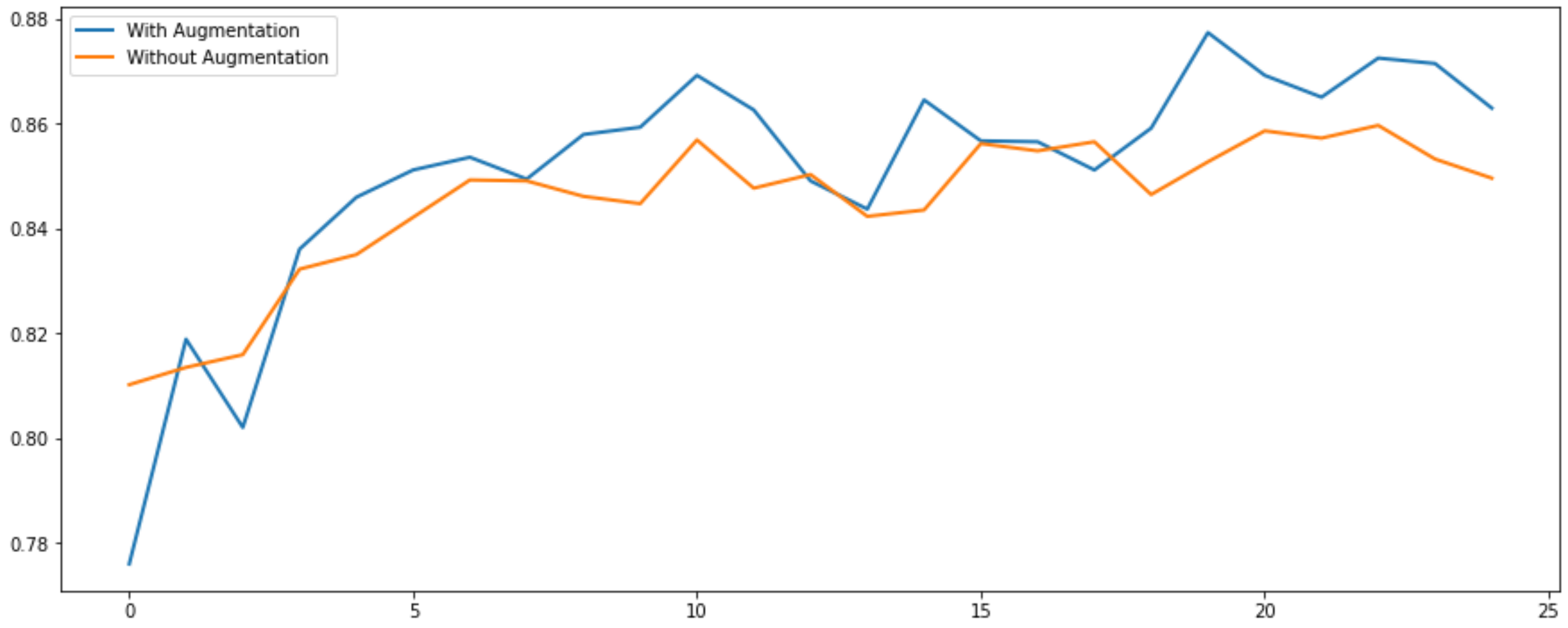

*Figure 8: Validation Accuracy of Data Augmentation Configuration*

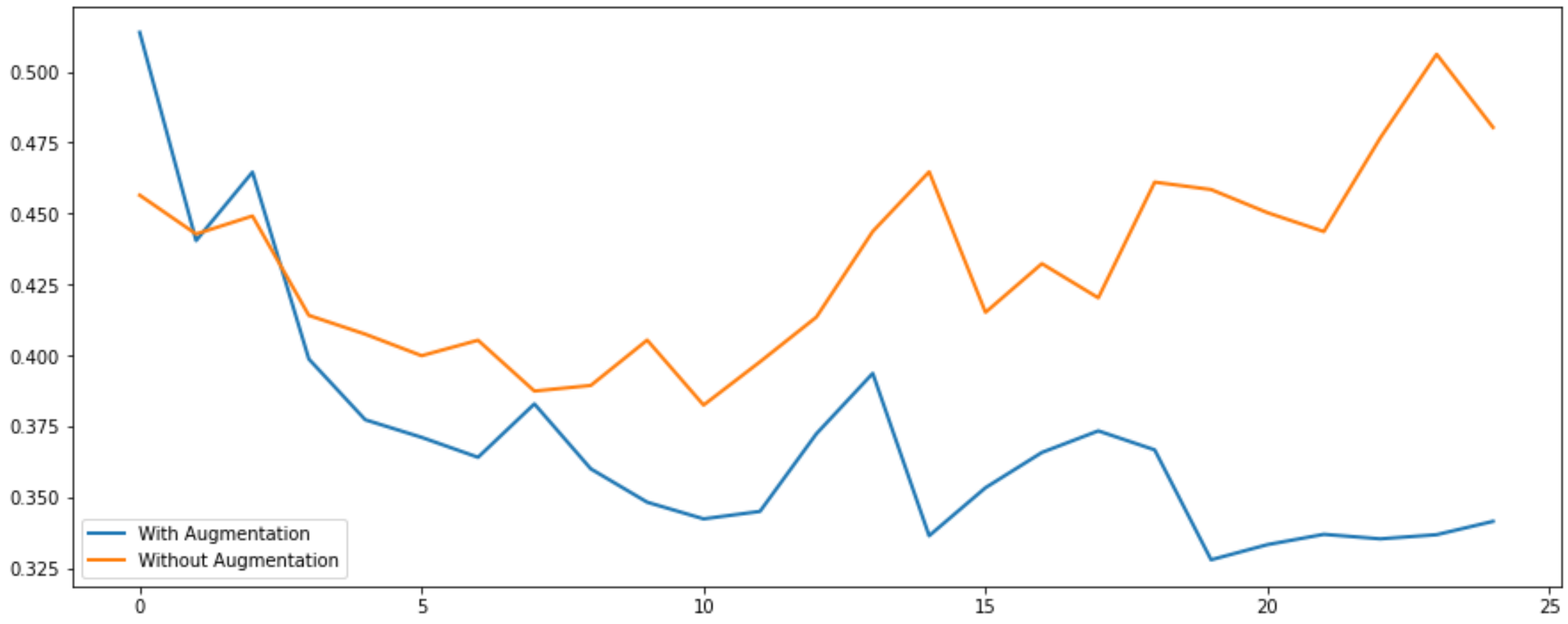

*Figure 9: Validation Loss of Data Augmentation Configuration*

Figure 5 shows the validation accuracy of a network with data augmentation consistently higher than a network with no augmentation. Figure 6 also shows that the validation loss of networks without data augmentation is increasing, while the validation loss of networks with data augmentation seems still decreasing. Thus, we will use data augmentation (random horizontal flip and random rotation 0.2) for our network.

#### 4.4.5 Learning Rate

Learning rate is a hyperparameter that determines how much to change model weights in response to the calculated error at each iteration. A learning rate that is too high will make the model become unstable, while a learning rate that is too low will be slower to reach the optimum performance. We will test three learning rates:

- Learning Rate 0.01
- Learning Rate 0.001 (the default of TensorFlow)
- Learning Rate 0.0005

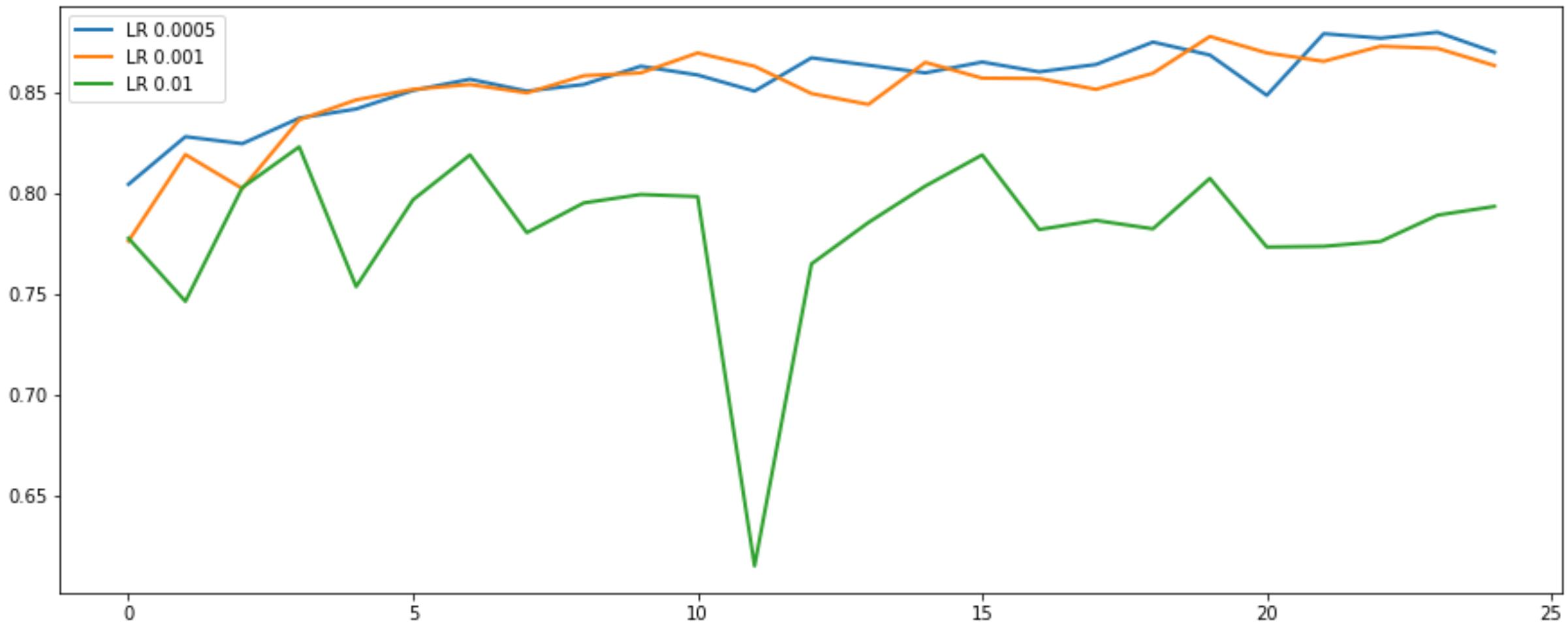

*Figure 10: Validation Accuracy of Multiple Learning Rate Configurations*

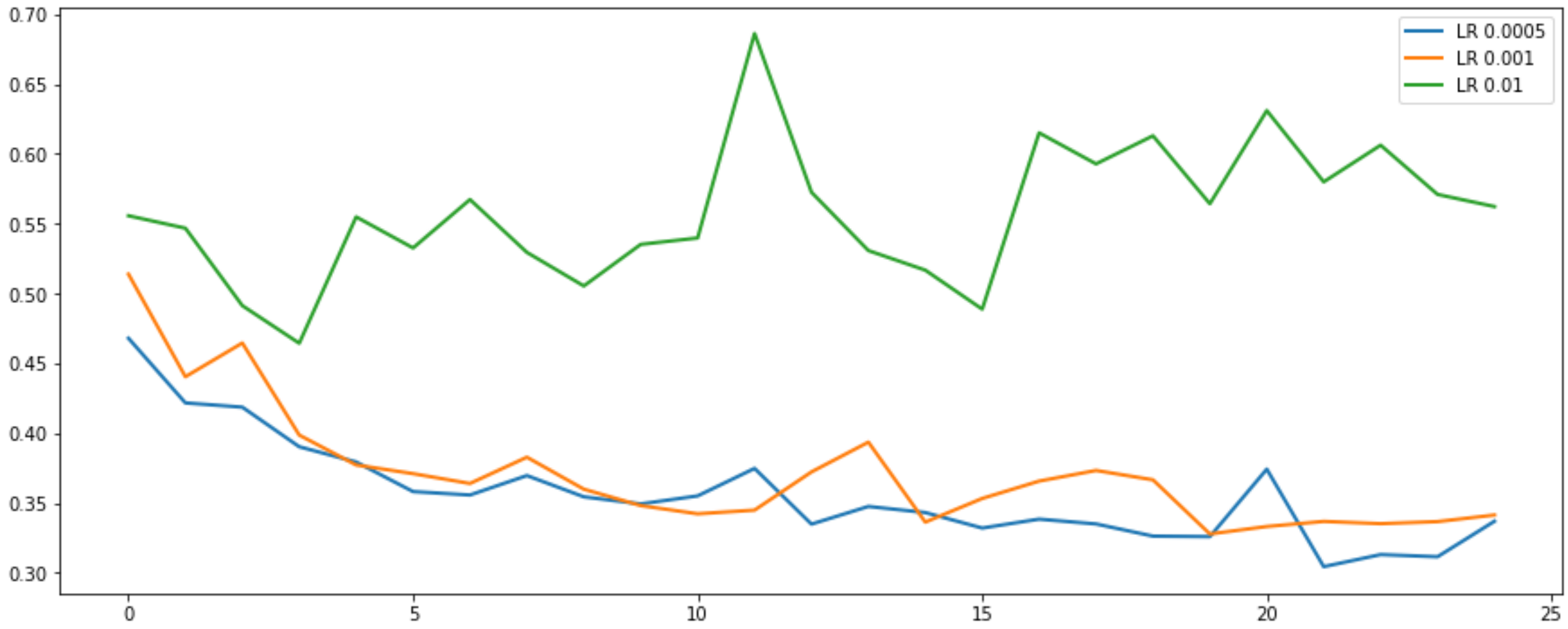

*Figure 11: Validation Loss of Multiple Learning Rate Configurations*

Figure 7 and Figure 8 show that the validation accuracy and validation loss of learning rate 0.01 differ much from the other learning rate. In other words, the learning rate of 0.01 is sub-optimal for our network. This leaves us two learning rates to choose from: 0.001 and 0.0005. It seems the performance is comparable between the two of them. We will try to test them again with bigger epochs in the next subsection.

### 4.4.6 Epochs

The number of epochs is influenced by and influences the learning rate and another hyperparameter. We will test two learning rates (0.001 and 0.0005) with a longer epoch.

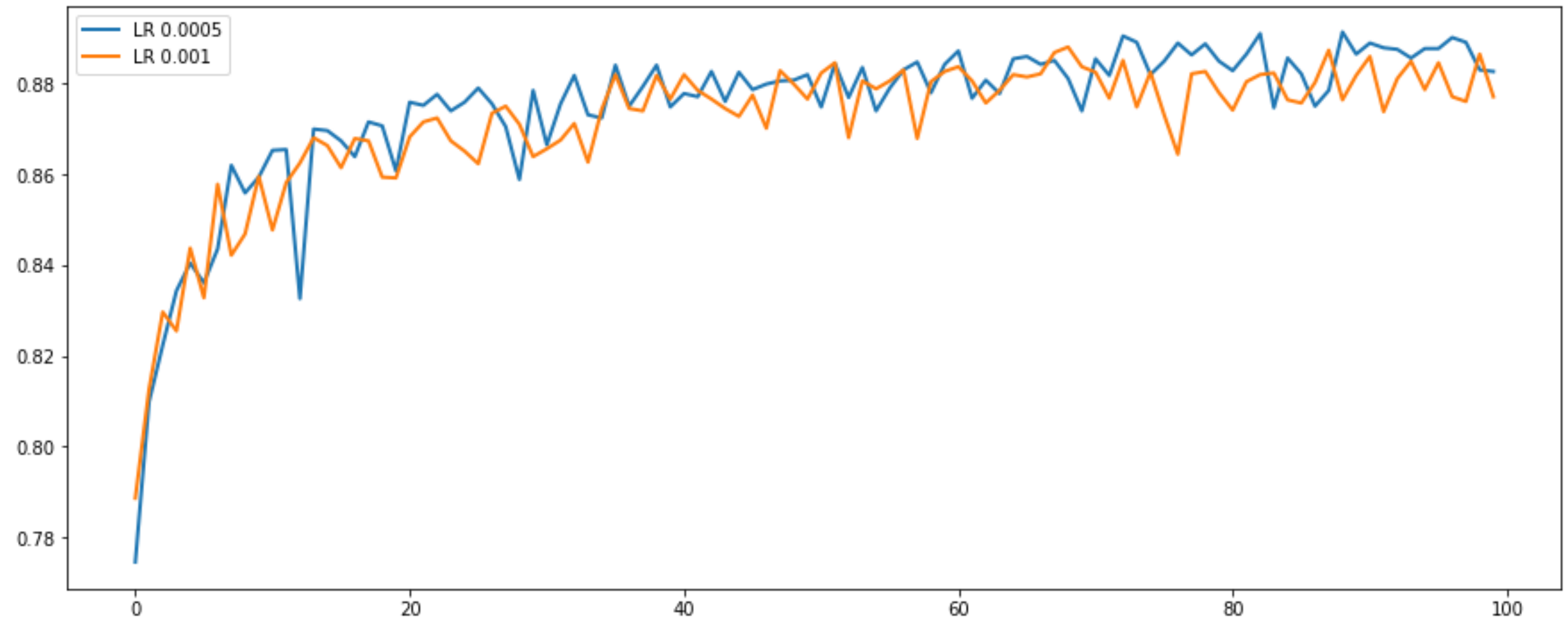

*Figure 12: Validation Accuracy of Learning Rates (100 Epochs)*

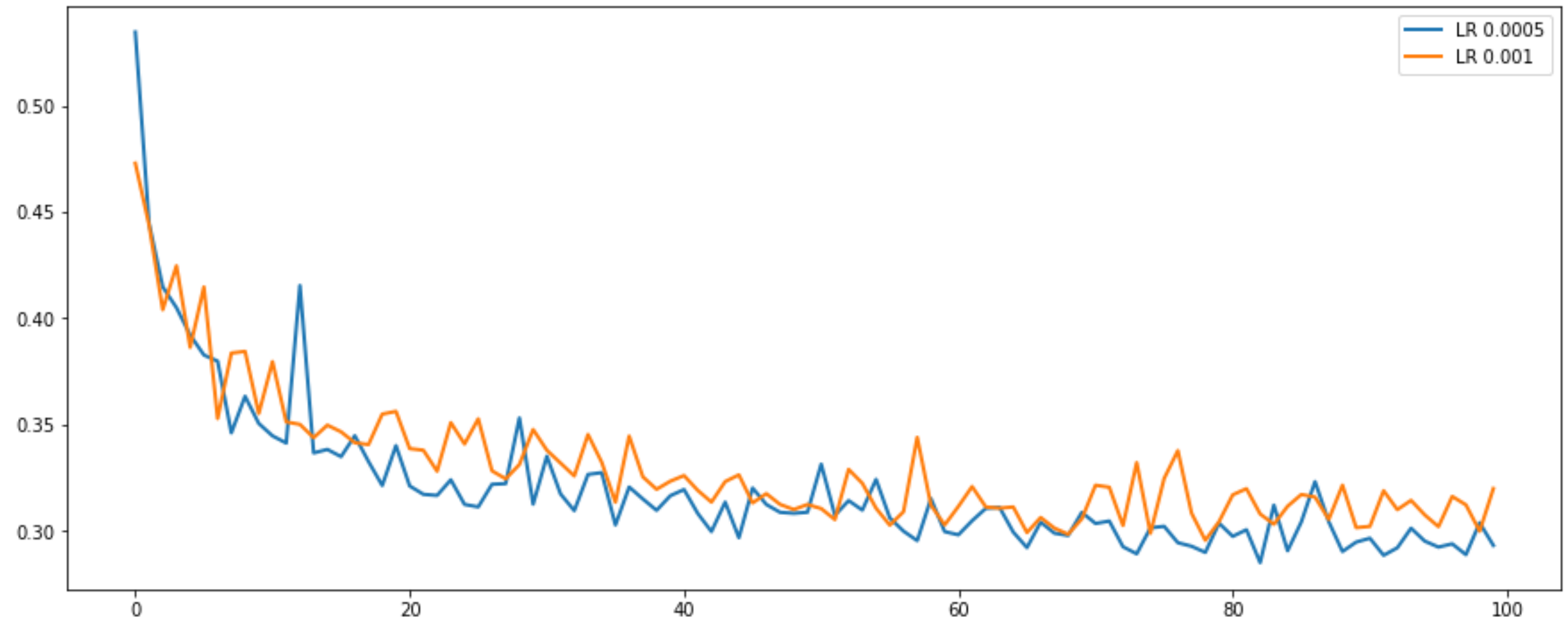

*Figure 13: Validation Loss of Learning Rates (100 Epochs)*

Figure 8 and Figure 9 show that the learning rate of 0.0005 is slightly better than the learning rate of 0.001. Thus, we will use a learning rate of 0.0005 for our networks.

### 4.4.7 Batch Size

Batch size is a hyperparameter that determines the number of data that will be passed through the network before updating the model. We will check three batch sizes:

- Batch size of 16
- Batch size of 32
- Batch size of 64

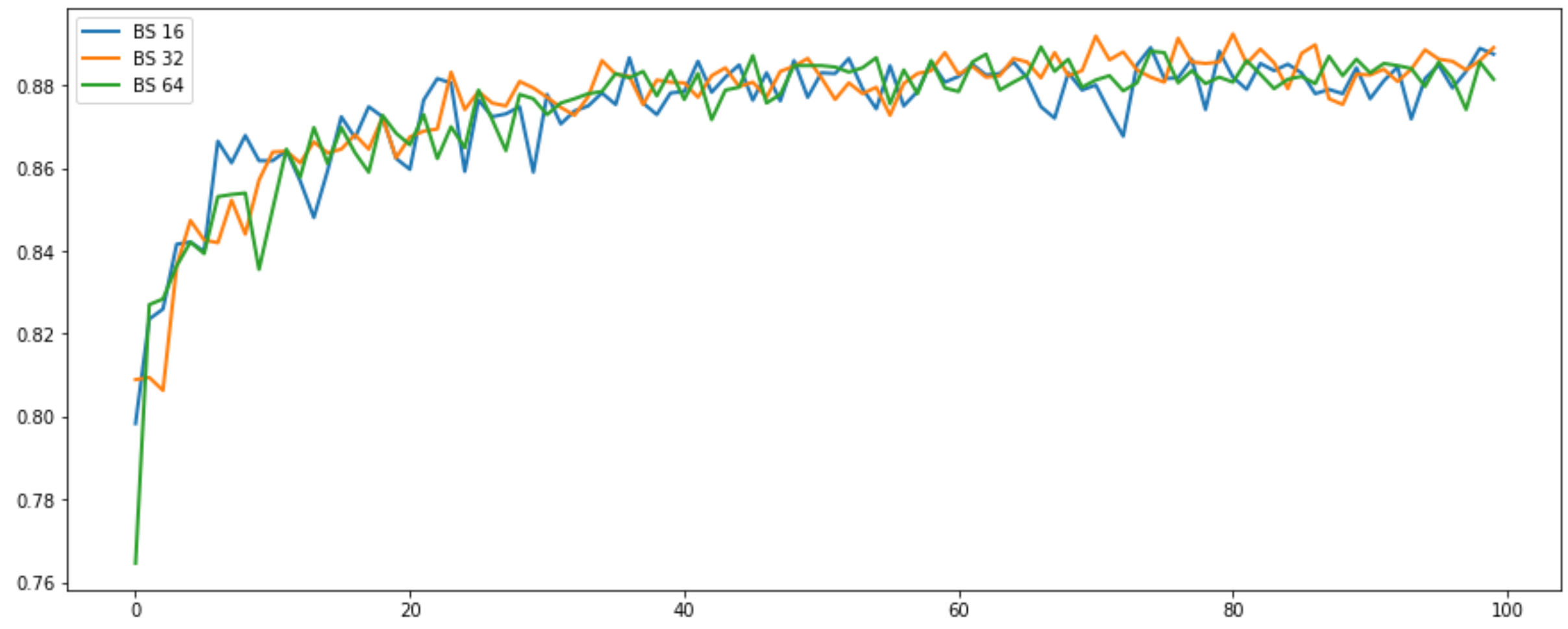

*Figure 14: Validation Accuracy of Multiple Batch Size Configurations*

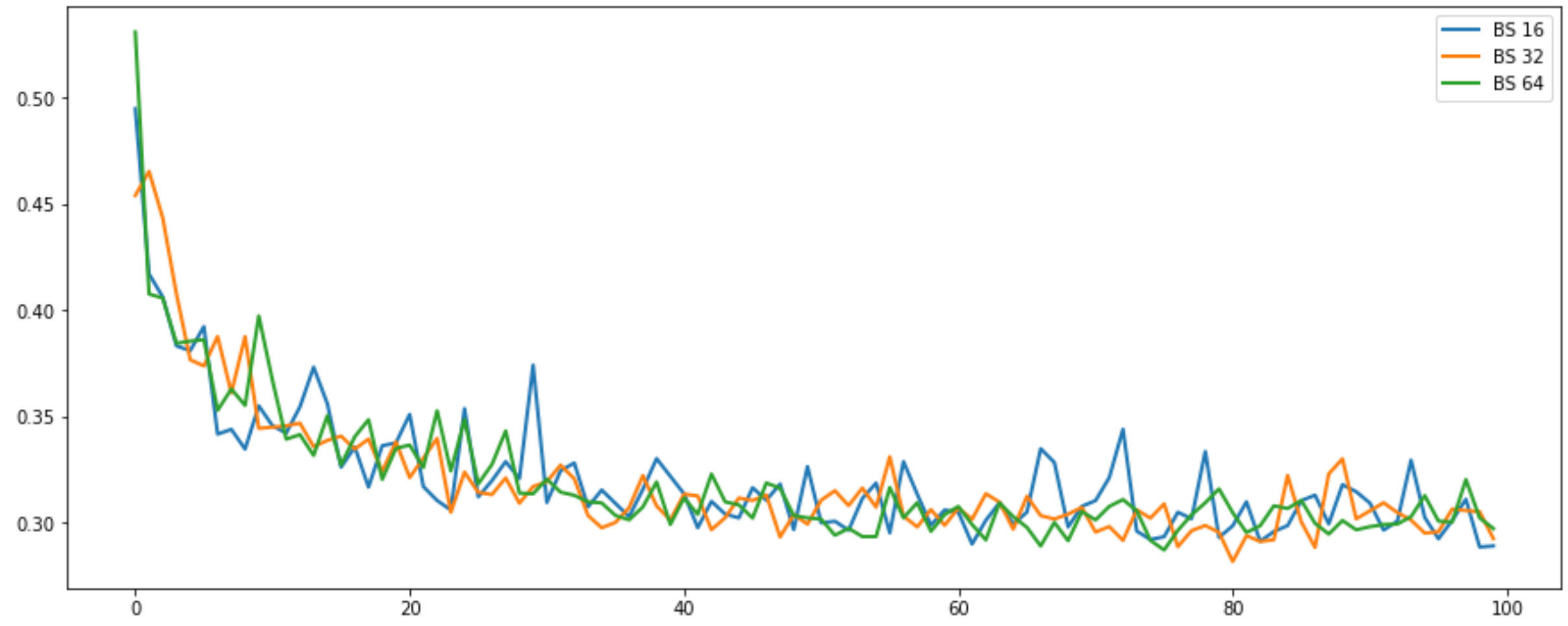

*Figure 15: Validation Loss of Multiple Batch Size Configurations*

Figure 11 and Figure 12 show that the performance of different batch size configurations is comparable. However, the training times are very different.

*Table 7: Batch Size and Training Time*

| **Batch Size** | **Training Time** |
|---|---|
| 16 | 2678 seconds |
| 32 | 1691 seconds |
| 64 | 1310 seconds |

Table 7 shows that as the batch size increases, the training time decreases. We will choose a batch size of 64 for our networks.

# CHAPTER 5

# RESULTS AND DISCUSSIONS

## 5.1 Introduction

This chapter will discuss the final results of our research. We will begin to describe the results of a single network in section 5.2, followed by the results of the ensemble in section 5.3.

## 5.2 Evaluation of Single Network

Following the implementation described in chapter 4, we test each combination of colour spaces and network architectures. Results indicated with * means that the colour is not further pre-processed by the built-in `preprocess_input` layer in TensorFlow.

The technical details are repeated here for clarity:

- Optimiser: Adam
- Top layers: Dense(512) + Dense(128) with Dropout(0.2)
- Data augmentation: random horizontal flip, random rotation of 0.2
- Learning rate: 0.0005
- Batch size: 64, epochs: 100
- Caching and prefetching enabled.

*Table 8: Single Network Classification Accuracy*

| | **RN50** | **RN101** | **RN152** | **VGG16** | **VGG19** | **DN121** | **DN169** | **DN201** | **Xception** |
|---|---|---|---|---|---|---|---|---|---|
| **RGB** | 0.8937 | 0.8836 | 0.8850 | 0.8451 | 0.8295 | 0.8407 | 0.8494 | 0.8651 | 0.8456 |
| **HLS*** | 0.8137 | 0.8270 | 0.8187 | 0.7779 | 0.7607 | 0.6705 | 0.6448 | 0.6506 | 0.6598 |
| **HLS** | 0.8277 | 0.8282 | 0.8282 | 0.7911 | 0.7864 | 0.7694 | 0.7930 | 0.7963 | 0.7576 |
| **HSV*** | 0.5759 | 0.5478 | 0.5617 | 0.7701 | 0.7560 | 0.7053 | 0.7332 | 0.7338 | 0.7209 |
| **HSV** | 0.5914 | 0.5653 | 0.5841 | 0.4297 | 0.4189 | 0.6030 | 0.6768 | 0.6923 | 0.5519 |
| **LAB*** | 0.8404 | 0.8435 | 0.8378 | 0.8321 | 0.8051 | 0.8098 | 0.8003 | 0.8244 | 0.8057 |
| **LAB** | 0.8637 | 0.8664 | 0.8498 | 0.8272 | 0.8154 | 0.8227 | 0.8288 | 0.8362 | 0.8532 |
| **LUV*** | **0.9015** | **0.9034** | **0.8918** | **0.8687** | **0.8499** | 0.7893 | 0.8090 | 0.8301 | 0.7061 |
| **LUV** | 0.8932 | 0.8885 | 0.8883 | 0.8678 | 0.8454 | 0.8611 | 0.8616 | 0.8724 | 0.8501 |
| **XYZ*** | 0.4802 | 0.6266 | 0.4965 | 0.7655 | 0.7595 | **0.8854** | **0.9074** | **0.9113** | **0.8894** |
| **XYZ** | 0.4575 | 0.6176 | 0.6070 | 0.2890 | 0.4338 | 0.4995 | 0.5073 | 0.5368 | 0.5809 |
| **YCrCb*** | 0.8840 | 0.8889 | 0.8696 | 0.8541 | 0.8449 | 0.7240 | 0.6926 | 0.7978 | 0.6165 |
| **YCrCb** | 0.8614 | 0.8609 | 0.8637 | 0.8519 | 0.8400 | 0.8301 | 0.8347 | 0.8513 | 0.8338 |
| **YDbDr*** | 0.4842 | 0.5582 | 0.5210 | 0.7567 | 0.7668 | 0.8234 | 0.8258 | 0.8204 | 0.8463 |
| **YDbDr** | 0.4849 | 0.5078 | 0.5010 | 0.2890 | 0.2890 | 0.5278 | 0.5967 | 0.6011 | 0.6091 |
| **YUV*** | 0.4486 | 0.5116 | 0.4515 | 0.7942 | 0.7977 | 0.8767 | 0.8868 | 0.8974 | 0.8807 |
| **YUV** | 0.5167 | 0.4894 | 0.4965 | 0.2890 | 0.3472 | 0.5061 | 0.5650 | 0.5740 | 0.5722 |

Table 8 shows the highest accuracy obtained from LUV and XYZ colour spaces. For ResNet and VGG family, LUV colour space returns the highest accuracy without using the pre-processing layer from the architecture. For DenseNet and Xception family, XYZ colour space returns the highest accuracy without using the pre-processing layer from the architecture.

*Table 9: Top 3 Highest Overall Accuracy for Single Network*

| **Configuration** | **Accuracy** |
|---|---|
| DenseNet201 + XYZ* | 0.9113 |
| DenseNet169 + XYZ* | 0.9074 |
| ResNet101 + LUV* | 0.9015 |

Table 9 shows the top 3 highest overall accuracy and its configuration. From the result presented in Table 8, the accuracy of converted images (non-RGB) is always higher than the original RGB images. The highest accuracy of RGB images is achieved by ResNet50 with the accuracy of 0.8937, while the highest overall accuracy is achieved by DenseNet201 with XYZ colour spaces with the accuracy of 0.9113.

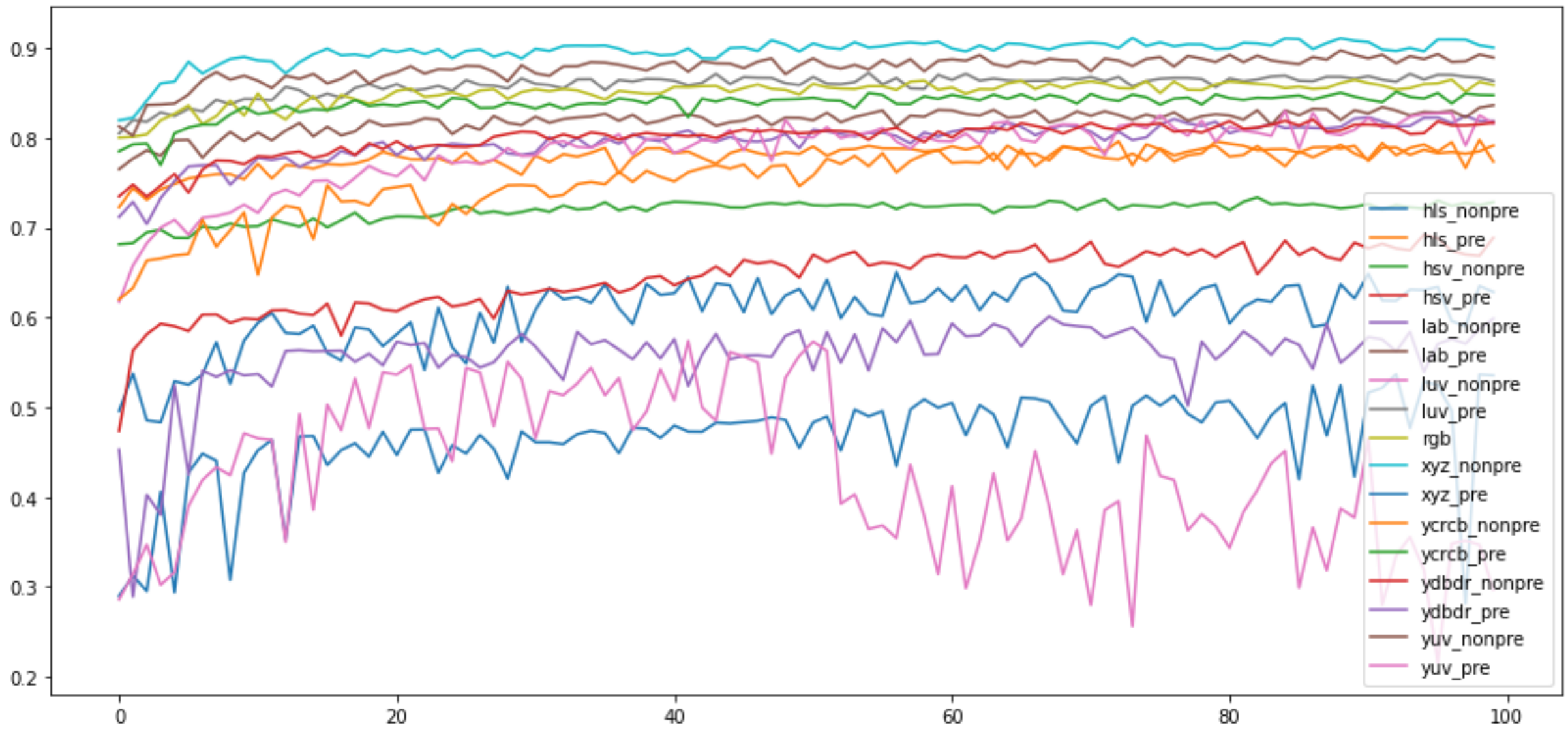

*Figure 16: Validation Accuracy of DenseNet201 with Multiple Colour Spaces*

Figure 16 shows the multiple colour spaces trained on DenseNet201. In some cases, colour space conversion increases the classification accuracy. However, some colour spaces decrease the classification accuracy, which may be caused by information loss caused by the colour space conversion.

For each architecture, the highest accuracy is obtained from the image that is not pre-processed using the preprocess_input layer. This makes sense because preprocess_input layer is usually made for RGB images. Thus, it is not suitable for converted images (non-RGB). For future studies, we recommend that researchers design an optimum preprocess_input layer for each colour space.

### 5.3 Evaluation of Ensemble Networks

Chapter 5.2 shows that the top validation accuracy for single network achieved by DenseNet201 with XYZ colour spaces. Thus, because of time and resource constraint, for this research we will focus on DenseNet family and XYZ colour spaces. We will also include Xception because it just happens to have the highest accuracy achieved by XYZ. In this section, we will test multiple ensembles configurations using DenseNet, Xception, XYZ, and RGB. The technical detail for ensemble networks is the same as the single network.

*Table 10: Validation Accuracy of Tested Ensemble Networks*

| **Configuration** | **Accuracy** |
|---|---|
| {Xception + DN121 + DN169 + DN201} XYZ* + {Xception + DN121 + DN169 + DN201} RGB | 0.9319 |
| {Xception + DN121 + DN169 + DN201} XYZ* | 0.9297 |
| {DN169 + DN201} XYZ* | 0.9258 |
| {Xception + DN169 + DN201} XYZ* | 0.9257 |
| {DN121 + DN169 + DN201} XYZ* | 0.9250 |
| {DN201 + Xception) XYZ* + {DN201 + Xception) RGB | 0.9241 |
| DN201 XYZ* + DN201 RGB | 0.9172 |
| {Xception + DN121 + DN169 + DN201} RGB | 0.9007 |
| {Xception + DN169 + DN201} RGB | 0.8974 |
| {DN121 + DN169 + DN201} RGB | 0.8908 |
| {DN169 + DN201} RGB | 0.8812 |

Table 10 shows the validation accuracy of multiple ensemble configurations. It is clear that combining two or more networks increases the accuracy. For example, the highest accuracy by a single model was achieved by DenseNet201 + XYZ* with an accuracy of 0.9113. Combining with an additional DenseNet201 network trained on RGB, the final accuracy increased to 0.9172. However, if we combine DenseNet201 + XYZ* with DenseNet169 + XYZ*, the final accuracy increased to 0.9258. Also, using RGB images only failed to surpass the XYZ* performances.

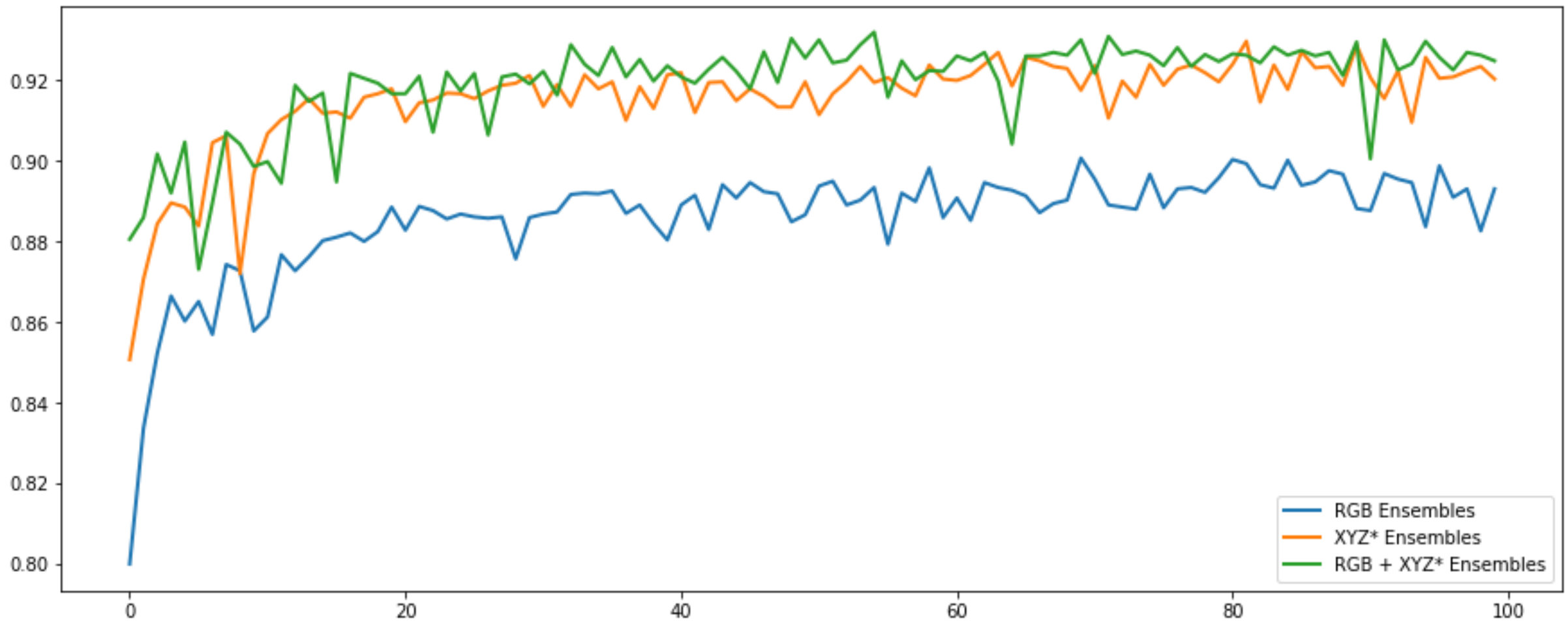

*Figure 17: Validation Accuracy of Ensembles of Four Networks with Multiple Colour Spaces*

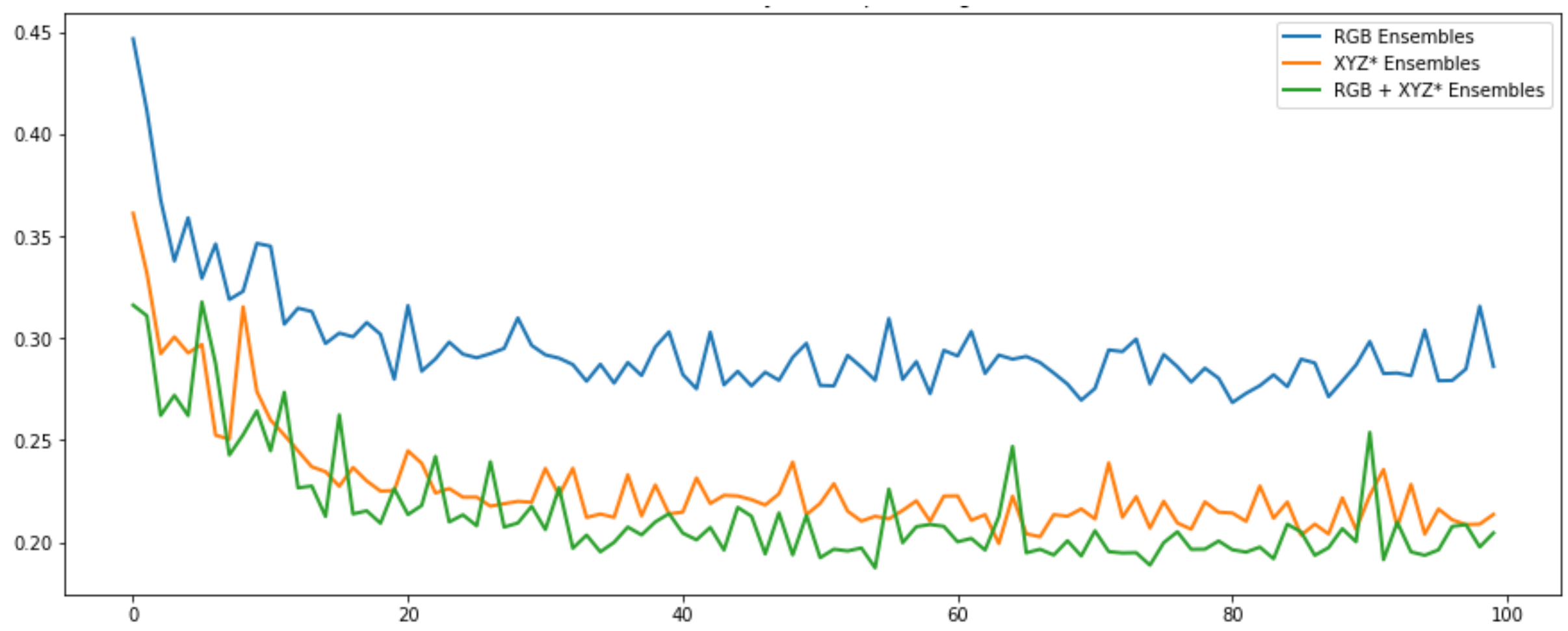

*Figure 18: Validation Loss of Ensembles of Four Networks with Multiple Colour Spaces*

Figure 17 shows the results of combining four networks (DenseNet121, DenseNet169, DenseNet201, and Xception) using three configurations: RGB only, XYZ* only, and a combination of XYZ and RGB (a total of eight networks). The figure shows that XYZ* colour spaces produce better results than the RGB images; it increases the accuracy from 0.9007 to 0.9297, which means a rise of 2.9 percent. Combining XYZ and RGB only increases the accuracy (Figure 17) and decreases the validation loss (Figure 18) by a marginal amount. This is not significant considering we achieve this by doubling the number of networks.

Using a smaller and fewer DenseNet, the accuracy increases becomes more significant. For example, using DenseNet169 and DenseNet201, shows that using XYZ* colour spaces increases the accuracy from 0.8812 to 0.9258, which means a rise of 4.46 percent. Looking at Table 10, there is something very interesting. Using fewer networks and parameters ({DN169 + DN201} XYZ*), we can achieve a similar or better results than using more networks and parameters ({Xception + DN121 + DN169 + DN201} RGB). This is achieved just by applying a colour space transformation (RGB to XYZ* in this case).

### 5.4 Summary

As described in this chapter, our findings can be summarised into three important points:

- Transforming image colour space into other colour spaces during pre-processing yields different results. Some colour spaces yield worse results, some colour spaces yield better results.
- By transforming colour space, we can use fewer networks and parameters to achieve results with similar or better results.

- Ensembles networks yield even better results. Thus, combining colour space transformation and ensembles produces superior results.

# CHAPTER 6

# CONCLUSION AND RECOMMENDATIONS

## 6.1 Introduction

This chapter is the final chapter of this thesis. Section 6.2 will discuss the result of this research. New contributions are described in Section 6.3. Future recommendations will be discussed in Section 6.4.

## 6.2 Discussion and Conclusion

The practice of colour space transformation in the pre-processing process has been used for some problems in the past. In this research, we conduct an experiment on whether colour space transformation will be helpful for astronomical problems.

The experimental result using individual network (DenseNet, ResNet, VGG, and Xception) and colour space transformation consistently showed a higher validation accuracy compared to a network that uses original RGB images. Ensembles network using multiple networks and colour spaces increases further the validation accuracy.

## 6.3 Contribution to Knowledge

Some contributions to knowledge from this research are:

- A pre-trained model that is trained on natural images can also be used for astronomical problems, which is considered to be a non-natural image.
- Colour space transformation can increase the classification accuracy of astronomical images (galaxy classification).
- Ensembles of multiple colour spaces and networks can further increase classification accuracy on astronomical images (galaxy classification).

## 6.4 Future Recommendations

Some configurations, tests, and experiments are not being done in this research due to the time and resource constraints (i.e., the author had limited access to the GPU computation power). Future works can include experimenting with more configurations and combinations to see if any interesting pattern emerges.

There are some ideas that could be tried in future works:

- Finding a more optimal hyperparameter for each architecture. In this research, we only find an optimal hyperparameter for one network and apply the same values to all networks. A different network may need a different hyperparameter to achieve its best performance.
- Using different layers in the network architecture. For example, we use MaxPool2D for the pooling layer in this research. However, we also can test whether AveragePooling2D could achieve better results.